\documentclass[aps,prd,nofootinbib,superscriptaddress,showpacs,showkeys,10pt]{revtex4}
\usepackage{txfonts}
\usepackage{graphicx}
\usepackage{dcolumn}
\usepackage{bm}
\usepackage{amssymb}
\usepackage{latexsym}
\usepackage[colorlinks, linkcolor=blue, citecolor=blue, urlcolor=blue]{hyperref}

\newcommand{\be}{\begin{equation}}
\newcommand{\ee}{\end{equation}}
\newcommand{\bq}{\begin{eqnarray}}
\newcommand{\eq}{\end{eqnarray}}

\begin{document}

\title{Comparison of dark energy models: A perspective from the latest observational data}

\author{Miao Li}
\email{mli@itp.ac.cn} \affiliation{Kavli Institute for Theoretical
Physics China, Chinese Academy of Sciences, Beijing 100190, China}
\affiliation{Key Laboratory of Frontiers in Theoretical Physics,
Institute of Theoretical Physics, Chinese Academy of Sciences,
Beijing 100190, China}

\author{Xiao-Dong Li}
\email{renzhe@mail.ustc.edu.cn} \affiliation{Interdisciplinary
Center for Theoretical Study, University of Science and Technology
of China, Hefei 230026, China} \affiliation{Key Laboratory of
Frontiers in Theoretical Physics, Institute of Theoretical Physics,
Chinese Academy of Sciences, Beijing 100190, China}

\author{Xin Zhang}
\email{zhangxin@mail.neu.edu.cn} \affiliation{Department of Physics,
College of Sciences, Northeastern University, Shenyang 110004,
China} \affiliation{Kavli Institute for Theoretical Physics China,
Chinese Academy of Sciences, Beijing 100190, China}

\begin{abstract}
In this paper, we compare some popular dark energy models under the
assumption of a flat universe by using the latest observational data
including the type Ia supernovae Constitution compilation, the
baryon acoustic oscillation measurement from the Sloan Digital Sky
Survey, the cosmic microwave background measurement given by the
seven-year Wilkinson Microwave Anisotropy Probe observations and the
determination of $H_0$ from the Hubble Space Telescope. Model
comparison statistics such as the Bayesian and Akaike information
criteria are applied to assess the worth of the models. These
statistics favor models that give a good fit with fewer parameters.
Based on this analysis, we find that the simplest cosmological
constant model that has only one free parameter is still preferred
by the current data. For other dynamical dark energy models, we find
that some of them, such as the $\alpha$ dark energy, constant $w$,
generalized Chaplygin gas, Chevalliear-Polarski-Linder
parametrization, and holographic dark energy models, can provide
good fits to the current data, and three of them, namely, the Ricci
dark energy, agegraphic dark energy, and Dvali-Gabadadze-Porrati
models, are clearly disfavored by the data.
\end{abstract}



\maketitle

\section{Introduction}\label{sec:intro}

Dark energy has become one of the most important issues of the
modern cosmology ever since the observations of type Ia supernovae
(SNe Ia) first indicated that the universe is undergoing an
accelerated expansion at the present stage \cite{Riess:1998cb}.
However, hitherto, we still know little about dark energy. The
limited information we know about dark energy includes: it causes
the cosmic acceleration; it accounts for two-thirds of the cosmic
energy density; it is gravitationally repulsive; it does not appear
to cluster in galaxies; and so on. Many cosmologists suspect that
the identity of dark energy is the cosmological constant that fits
the observational data well. While, one also has reason to dislike
the cosmological constant since it always suffers from the
theoretical problems such as the ``fine-tuning'' and ``cosmic
coincidence'' puzzles \cite{DErev}. The fine-tuning problem, also
known as the ``old cosmological constant problem,'' is motivated by
the enormous discrepancy between the theoretical prediction for the
cosmological constant and its measured value. The so-called ``new
cosmological constant problem,'' namely, the cosmic coincidence
problem, questions why we just live in an era when the densities of
dark energy and matter are almost equal, which also indicates that
the cosmological constant scenario may be incomplete. Thus, a
variety of proposals for dark energy have emerged.

The possibility that dark energy is dynamical, for example, in a
form of some light scalar field \cite{quintessence}, has been
explored by cosmologists for a long time. A basic way to explore
such a dynamical dark energy model in light of observational data is
to parameterize dark energy by an equation-of-state parameter $w$,
relating the dark energy pressure $p$ to its density $\rho$ via
$p=w\rho$. In general, this parameter $w$ is time variable. The most
commonly used forms of $w(a)$ involve the constant equation of
state, $w={\rm const.}$, and the Chevalliear-Polarski-Linder form
\cite{CPL}, $w(a)=w_0+(1-a)w_a$, where $w_0$ and $w_a$ parameterize
the present-day value of $w$ and the first derivative. There are
also many other dynamical dark energy models which stem from
different aspects of new physics. For example, the ``holographic
dark energy'' models \cite{hde,holoext,oldade,ade,ageext,rde,rdeext}
arise from the holographic principle of quantum gravity theory, and
the Chaplygin gas models \cite{cg,gcg,ngcg} are motivated by brane
world scenarios and may be able to unify dark matter and dark
energy. In addition, there is also significant interest in
modifications to general relativity, in the context of explaining
the acceleration of the universe. The Dvali-Gabadadze-Porrati models
\cite{DGP,DGPext,alphaDE} arise from a class of brane-related
theories in which gravity leaks out into the bulk at large
distances, leading to the accelerated expansion of the universe.

In the face of so many competing dark energy candidates, it is
important to find an effective way to decide which one is right, or
at least, which one is most favored by the observational data.
Although the accumulation of the current observational data has
opened a robust window for constraining the parameter space of dark
energy models, the model filtration is still a difficult mission
owing to the accuracy of current data as well as the complication
caused by different parameter numbers of various dark energy models.
In this paper, we make an effort to assess some popular dark energy
models in light of the latest observational data, including the
Constitution SN data \cite{SN09} and other cosmological probes such
as the distance information measured by Wilkinson Microwave
Anisotropy Probe (WMAP) \cite{WMAP7}, and the Baryon Acoustic
Oscillations (BAO) \cite{BAO,BAO2}. To make a comparison for various
dark energy models with different numbers of parameters and decide
on the model preferred by the current data, following
Ref.~\cite{Davis:2007na}, we apply model comparison statistics such
as the Bayesian information criterion (BIC) \cite{BIC} and the
Akaike information criterion (AIC) \cite{AIC} in our analysis.

This paper is organized as follows. In Sec. \ref{sec:method}, we
discuss the information criteria in the context of dark energy model
selection. In Sec. \ref{sec:data}, we give details of the
observational data sets used. Section \ref{sec:model} describes nine
popular dark energy models and assesses which one is preferred by
the current data. The results are discussed in Sec. \ref{sec:concl}.

\section{Methodology}\label{sec:method}

In this work we employ the $\chi^2$ statistics. For a physical
quantity $\xi$ with experimentally measured value $\xi_{obs}$,
standard deviation $\sigma_{\xi}$, and theoretically predicted value
$\xi_{th}$, the $\chi^2$ value is given by \be \label{eq:chi2_xi}
\chi_{\xi}^2=\frac{\left(\xi_{th}-\xi_{obs}\right)^2}{\sigma_{\xi}^2}.
\ee The total $\chi^2$ is the sum of all $\chi_{\xi}^2$s, i.e., \be
\label{eq:chi2} \chi^2=\sum_{\xi}\chi_{\xi}^2. \ee The observational
data we use in this paper include the Constitution SN Ia sample, the
Cosmic Microwave Background (CMB) measurement given by the
seven-year WMAP observations, the BAO measurement from the Sloan
Digital Sky Survey (SDSS), and the measurement of $H_0$ from the
Hubble Space Telescope (HST) \cite{H0}.

However, the $\chi^2$ statistic alone cannot provide effective way
to make a comparison between competing models since this method is
based on the assumption that the underlying model is the correct
one. The $\chi^2$ statistics is good at finding the best-fit values
of parameters but is insufficient for deciding whether the model
itself is the best one. Since in general a model with more
parameters tends to give a lower $\chi^2_{min}$, it is unwise to
compare different models by simply considering $\chi^2_{min}$ with
likelihood contours or best-fit parameters. Instead, one may employ
the information criteria (IC) to assess different models, which is
also based on a likelihood method. In this paper, we use the BIC
\cite{BIC} and AIC \cite{AIC} as model selection criteria. According
to these criteria, models that give a good fit with fewer parameters
will be more favored. So, these criteria embody the principle of
Occam's razor, ``entities must not be multiplied beyond necessity.''
The applications of the BIC and AIC in a cosmological context can be
found in, e.g., Refs. \cite{Liddle:2004nh,cosmologyIC}.

The BIC, also known as the Schwarz information criterion \cite{BIC},
is given by
\begin{equation}
{\rm BIC}=-2\ln{\cal L}_{max}+k\ln N,
\end{equation}
where ${\cal L}_{max}$ is the maximum likelihood, $k$ is the number
of parameters, and $N$ is the number of data points used in the fit.
Note that for Gaussian errors, $\chi_{min}^2=-2\ln{\cal L}_{max}$,
and the difference in BIC can be simplified to $\Delta{\rm
BIC}=\Delta\chi_{min}^2+\Delta k\ln N$. A difference in BIC
($\Delta{\rm BIC}$) of 2 is considered positive evidence against the
model with the higher BIC, while a $\Delta{\rm BIC}$ of 6 is
considered strong evidence.

The AIC \cite{AIC} is defined as
\begin{equation}
{\rm AIC}=-2\ln{\cal L}_{max}+2k,
\end{equation}
which gives results similar to the BIC approach, but it should be
pointed out that the AIC is more lenient than BIC on models with
extra parameters for any likely data set $\ln N>2$. 
Also, in this case, the absolute value of the criterion is not of
interest, only the relative value between different models,
$\Delta{\rm AIC}=\Delta\chi_{min}^2+2\Delta k$, is useful. As
mentioned in Ref.~\cite{Liddle:2007fy}, there is a version of the
AIC corrected for small sample sizes, ${\rm AIC}_c={\rm
AIC}+2k(k-1)/(N-k-1)$, which is important for $N/k\lesssim 40$.
Obviously, in our case, this correction is negligible.

It should be noted that the information criteria alone can at most
say that a more complex model is not necessary to explain the
current data, since a poor information criterion result might arise
from the fact that the current data are too poor to constrain the
extra parameters in this complex model, and it might become
preferred with improved data. Actually, this is just the current
situation for dynamical dark energy models.

A more sophisticated method for model selection is provided by the
Bayesian evidence which does not simply count parameters, but
considers how much the allowed volume in data space increases due to
the addition of extra parameters, as well as any correlations
between the parameters. So, the Bayesian evidence requires an
integral of the likelihood over the whole model parameter space,
which may be lengthy to calculate, but avoids the approximations
used in the information criteria and also permits the use of prior
information if required. This method has been applied in a variety
of cosmological contexts; see, e.g.,
Refs.~\cite{Saini:2003wq,Liddle:2006kn,holomodels09}. Information
criteria require no assumptions for the prior or the metric on the
space of model parameters. In this paper, we will use the first
approximation provided by the information criteria without
calculating the full Bayesian evidence. This simpler version is
sufficient for our purpose.

\section{Current Observational Data}\label{sec:data}

In order to test the different dark energy models, we have used the
observational data currently available. In this section, we describe
how we use these data.

\subsection{Type Ia supernovae}

Up to now, SNe Ia provide the most direct indication of the
accelerated expansion of the universe. It is commonly believed that
these SNe Ia all have the same intrinsic luminosity, and thus they
are used as ``standard candles.'' Therefore, measuring both their
redshift and their apparent peak flux gives a direct measurement of
their luminosity distance $d_L$ as a function of redshift $z$. The
function $d_L(z)$ encodes the expansion history of the universe so
that by which the information of dark energy can be extracted.

In this paper, for the SN Ia data, we use the Constitution sample
including 397 data that are given in terms of the distance modulus
$\mu_{ obs}(z_i)$ compiled in Table 1 of Ref.~\cite{SN09}. The
theoretical distance modulus is defined as
\begin{equation}
\mu_{th}(z_i)\equiv 5 \log_{10} {D_L(z_i)} +\mu_0,
\end{equation}
where $\mu_0\equiv 42.38-5\log_{10}h$ with $h$ the Hubble constant
$H_0$ in units of 100 km/s/Mpc, and the Hubble-free luminosity
distance $D_L=H_0d_L$ is
\begin{equation}
D_L(z)={1+z\over \sqrt{|\Omega_{k}|}}\textrm{sinn}\Big(
\sqrt{|\Omega_{k}|}\int_0^z{dz'\over E(z')} \Big),
\end{equation}
where $E(z)\equiv H(z)/H_0$, $\Omega_{k}$ is the fractional
curvature density at $z=0$, and
\begin{displaymath}
{\textrm{sinn}\left(\sqrt{|\Omega_{k}|}x\right)\over
\sqrt{|\Omega_{k}|}} = \left\{
\begin{array}{ll}
{\textrm{sin}(\sqrt{|\Omega_{k}|}x)/
\sqrt{|\Omega_{k}|}}, & \textrm{if $\Omega_{k}<0$ (or $k=+1$)},\\
x, & \textrm{if $\Omega_{k}=0$ (or $k=0$)},\\
{\textrm{sinh}(\sqrt{|\Omega_{k}|}x)/
 \sqrt{|\Omega_{k}|}}, &
\textrm{if $\Omega_{k}>0$ (or $k=-1$)}.
\end{array} \right.
\end{displaymath}

The $\chi^2$ for the SN data is
\begin{equation}
\chi^2_{SN}({\bm\theta})=\sum\limits_{i=1}^{397}{[\mu_{obs}(z_i)-\mu_{th}(z_i;{\bm\theta})]^2\over
\sigma_i^2},\label{ochisn}
\end{equation}
where $\mu_{obs}(z_i)$ and $\sigma_i$ are the observed value and the
corresponding 1$\sigma$ error of distance modulus for each
supernova, respectively, and ${\bm\theta}$ denotes the model
parameters.

What should be mentioned is that since the absolute magnitude of a
supernova is unknown, the degeneracy between the Hubble constant and
the absolute magnitude implies that one cannot quote constraints on
either one. Thus the nuisance parameter $H_0$ in SN data is not the
observed Hubble constant and is different from that in the BAO and
CMB data. Therefore we should analytically marginalize over $H_0$ in
the SN data.

Following Refs.~\cite{Nesseris:2005ur}, the minimization with
respect to $\mu_0$ can be made trivially by expanding the $\chi^2$
of Eq. (\ref{ochisn}) with respect to $\mu_0$ as \be
\chi^2_{SN}({\bm\theta})=A({\bm\theta})-2\mu_0
B({\bm\theta})+\mu_0^2 C, \ee where
 \be A({\bm\theta})=\sum\limits_{i}{[\mu_{obs}(z_i)-\mu_{th}(z_i;\mu_0=0,{\bm\theta})]^2\over
\sigma_i^2}, \ee \be
B({\bm\theta})=\sum\limits_{i}{\mu_{obs}(z_i)-\mu_{th}(z_i;\mu_0=0,{\bm\theta})\over
\sigma_i^2}, \ee \be C=\sum\limits_{i}{1\over \sigma_i^2}. \ee
Evidently, Eq. (\ref{ochisn}) has a minimum for $\mu_0=B/C$ at \be
\tilde{\chi}^2_{
SN}({\bm\theta})=A({\bm\theta})-{B({\bm\theta})^2\over
C}.\label{tchi2sn} \ee Since $\chi^2_{SN, min}=\tilde{\chi}^2_{SN,
min}$, instead minimizing $\chi^2_{SN}$ we will minimize
$\tilde{\chi}^2_{SN}$ which is independent of the nuisance parameter
$\mu_0$.

\subsection{Baryon Acoustic Oscillations}

Aside from the SN data, the other external astrophysical results
that we shall use in this paper for the joint cosmological analysis
are the BAO and CMB data. First, we describe how we use the BAO
data. The BAO is a powerful probe of dark energy, since it can be
used to measure not only the angular diameter distance $D_A(z)$
through the clustering perpendicular to the line of sight, but also
the expansion rate of the universe $H(z)$ through the clustering
along the line of sight. However, the current data are not accurate
enough to allow us to extract $D_A(z)$ and $H(z)$ separately.
Actually, the BAO currently can barely be measured in the
spherically-averaged correlation function \cite{Okumura:2007br}.

The spherical average gives us the following effective distance
measure \cite{bao05}
\begin{equation}
 D_V(z) \equiv \left[(1+z)^2D_A^2(z)\frac{z}{H(z)}\right]^{1/3},
\end{equation}
where $D_A(z)$ is the proper (not comoving) angular diameter
distance,
\begin{equation}
 D_A(z) = \frac{1}{H_0}\frac{\textrm{sinn}(\sqrt{|\Omega_{k}|}H_0\int_0^z\frac{dz'}{H(z')})/ \sqrt{|\Omega_{k}|}}
 {(1+z)}.
\label{eq:da}
\end{equation}

The BAO data from the spectroscopic SDSS Data Release 7 (DR7) galaxy
sample \cite{BAO2} give $D_V(z=0.35)/D_V(z=0.2)=1.736\pm 0.065$.
Thus, the $\chi^2$ for BAO data is,
\begin{equation}
\chi^2_{BAO}=\left(\frac{D_V(z=0.35)/D_V(z=0.2)-1.736}{0.065}\right)^2.
\end{equation}

\subsection{Cosmic Microwave Background}

The CMB is sensitive to the distance to the decoupling epoch via the
locations of peaks and troughs of the acoustic oscillations. In this
paper, we employ the ``WMAP distance priors'' given by the
seven-year WMAP observations \cite{WMAP7}. This includes the
``acoustic scale'' $l_A$, the ``shift parameter'' $R$, and the
redshift of the decoupling epoch of photons $z_*$.

The acoustic scale $l_A$ describes the distance ratio
$D_A(z_*)/r_s(z_*)$, defined as
\begin{equation}
\label{ladefeq} l_A\equiv (1+z_*){\pi D_A(z_*)\over r_s(z_*)},
\end{equation}
where a factor of $(1+z_*)$ arises because $D_A(z_*)$ is the proper
angular diameter distance, whereas $r_s(z_*)$ is the comoving sound
horizon at $z_*$. The fitting formula of $r_s(z)$ is given by
\begin{equation}
r_s(z)=\frac{1} {\sqrt{3}}  \int_0^{1/(1+z)}  \frac{ da } { a^2H(a)
\sqrt{1+(3\Omega_{b}/4\Omega_{\gamma})a} },
\end{equation}
where $\Omega_{b}$ and $\Omega_{r}$ are the present-day baryon and
photon density parameters, respectively. In this paper, we fix
$\Omega_{\gamma}=2.469\times10^{-5}h^{-2}$ (for $T_{cmb}=2.725$ K)
and $\Omega_{b}=0.022765 h^{-2}$, which are the best-fit values
given by the seven-year WMAP observations \cite{WMAP7}. We use the
fitting function of $z_*$ proposed by Hu and Sugiyama
\cite{Hu:1995en}:
\begin{equation}
\label{zstareq} z_*=1048[1+0.00124(\Omega_b
h^2)^{-0.738}][1+g_1(\Omega_m h^2)^{g_2}],
\end{equation}
where
\begin{equation}
g_1=\frac{0.0783(\Omega_b h^2)^{-0.238}}{1+39.5(\Omega_b
h^2)^{0.763}},\quad g_2=\frac{0.560}{1+21.1(\Omega_b h^2)^{1.81}}.
\end{equation}

The shift parameter $R$ is responsible for the distance ratio
$D_A(z_*)/H^{-1}(z_*)$, given by \cite{Bond97}
\begin{equation}
\label{shift} R(z_*)\equiv \sqrt{\Omega_m H_0^2}(1+z_*)D_A(z_*).
\end{equation}
Actually, this quantity is different from $D_A(z_*)/H^{-1}(z_*)$ by
a factor of $\sqrt{1+z_*}$, and also ignores the contributions from
radiation, curvature, or dark energy to $H(z_*)$. Nevertheless, we
still use $R$ to follow the convention in the literature.

Following Ref.~\cite{WMAP7}, we use the prescription for using the
WMAP distance priors. Thus, the $\chi^2$ for the CMB data is
\begin{equation}
\chi_{CMB}^2=(x^{th}_i-x^{obs}_i)(C^{-1})_{ij}(x^{th}_j-x^{obs}_j),\label{chicmb}
\end{equation}
where $x_i=(l_A, R, z_*)$ is a vector, and $(C^{-1})_{ij}$ is the
inverse covariance matrix. The seven-year WMAP observations
\cite{WMAP7} give the maximum likelihood values: $l_A(z_*)=302.09$,
$R(z_*)=1.725$, and $z_*=1091.3$. The inverse covariance matrix is
also given in Ref.~\cite{WMAP7}:
\begin{equation}
(C^{-1})=\left(
  \begin{array}{ccc}
    2.305 & 29.698 & -1.333 \\
    29.698& 6825.27 & -113.180 \\
    -1.333& -113.180 &  3.414 \\
  \end{array}
\right).
\end{equation}

\subsection{Hubble constant}
In this paper we also use the prior on the present-day Hubble
constant, $H_0=74.2\pm 3.6$km/s/Mpc \cite{H0}. In Ref.~\cite{H0},
the authors obtain this measurement result of $H_0$ from the
magnitude-redshift relation of 240 low-$z$ type Ia supernovae at
$z<0.1$. The absolute magnitudes of these supernovae are calibrated
by using new observations from HST of 240 Cepheid variables in six
local type Ia supernovae host galaxies and the maser galaxy NGC
4258. It is remarkable that this Gaussian prior on $H_0$ has also
been used in the analysis of WMAP 7-year observational data
\cite{WMAP7}. The $\chi^2$ function for the Hubble constant is
\begin{equation}
\chi^2_{h}=\left(\frac{h-0.742}{0.036} \right)^2.
\end{equation}

\subsection{Combining the constraints}

Since the SN, BAO, CMB and $H_0$ are effectively independent
measurements, we can combine our results by simply adding together
the $\chi^2$ functions. Thus, we have
\begin{equation}
\chi^2=\tilde{\chi}^2_{SN}+\chi^2_{BAO}+\chi^2_{CMB}+\chi^2_{h}.
\end{equation}
Note that $\tilde{\chi}^2_{SN}$ and $\chi^2_{BAO}$ are free of $h$,
while $\chi^2_{CMB}$ and $\chi^2_{h}$ are still relevant to $h$.

\section{Dark Energy Models}\label{sec:model}

In the standard homogeneous and isotropic Friedmann-Robertson-Walker
(FRW) universe, the Friedmann equation is expressed as
\begin{equation}
3M_{Pl}^2H^2=\rho-{3M_{Pl}^2 k\over a^2},\label{fkeq}
\end{equation}
where $M_{Pl}\equiv 1/\sqrt{8\pi G}$ is the reduced Planck mass,
$\rho$ is the total energy density containing contributions from
cold dark matter, baryons, radiations, and dark energy, and $k$
describes the spatial geometry of the universe. Usually, the
properties of dark energy, such as its equation of state parameter
$w$, are degenerate with the spatial curvature of the universe
$\Omega_k$. Thus, although in principle one should include
$\Omega_k$ as an additional parameter when fitting dark energy
models in light of observational data, the current data are not
accurate enough to distinguish between $w(z)$ and $\Omega_k$, owing
to the degeneracy of them. On the other hand, it is well known that
most inflation models in which the inflationary periods last for
much longer than 60 $e$-folds predict a spatially flat universe,
$\Omega_k\sim 10^{-5}$. Actually, the inflation theory has become a
paradigm in the modern cosmology and it has received strong support
from the CMB observations. Under such circumstances, therefore, in
this paper we shall use a ``strong inflation prior,'' imposing a
flatness prior, and explore dark energy models in the context of
such inflation models.

In a spatially flat FRW universe $(\Omega_k=0)$, the Friedmann
equation (\ref{fkeq}) reduces to \be \label{Fried}
3M_{Pl}^2H^2=\rho_{m}(1+z)^3+\rho_{r}(1+z)^4+\rho_{de}(0)f(z),\ee
where $\rho_m$, $\rho_r$ and $\rho_{de}(0)$ are the present-day
densities of dust matter, radiation and dark energy, respectively,
and $f(z)\equiv \rho_{de}(z)/\rho_{de}(0)$ is given by the specific
dark energy models. This equation is usually rewritten as \be
\label{Ez1} E(z)\equiv H(z)/H_{0}
=\left[\Omega_{m}(1+z)^3+\Omega_{r}(1+z)^4+(1-\Omega_{m}-\Omega_{r})f(z)\right]^{1/2}.
\ee Note that the radiation density parameter $\Omega_r$ is the sum
of the photons and relativistic neutrinos \cite{WMAP5},
\begin{equation}
\Omega_r=\Omega_\gamma(1 + 0.2271N_{eff}),
\end{equation}
where $N_{eff}$ is the effective number of neutrino species, and in
this paper we take its standard value 3.04 \cite{WMAP7}. In some
cases, the evolution of the dark energy density parameter
$\Omega_{de}(z)=\rho_{de}(z)/(3M_{Pl}H^2)$ is determined by a
differential equation, and thus one should express the Friedmann
equation as \be \label{Ez2}
E(z)=\left(\Omega_{m}(1+z)^3+\Omega_{r}(1+z)^4\over
1-\Omega_{de}(z)\right)^{1/2}. \ee

In what follows, we choose nine popular dark energy models and
examine whether they are consistent with the data currently
available to us. We divide these models into five classes:
\begin{enumerate}
\item Cosmological constant model.
\item Dark energy models with equation of state parameterized.
\item Chaplygin gas models.
\item Holographic dark energy models.
\item Dvali-Gabadadze-Porrati (DGP) brane world and related models
\end{enumerate}

It should be mentioned that the dark energy models discussed in this
paper involve those invoke a variation in the equations governing
gravity, say, the DGP model. Actually, in some cases the two are
interchangeable descriptions of a single theory, albeit they might
affect the structure growth in different manners (see, e.g.,
Refs.~\cite{Zhang:2007nk,Jain:2007yk}). In this paper, we ignore the
exiguous difference between them, and call them uniformly the ``dark
energy models.'' The models discussed and the parameters that
describe each model are summarized in Table \ref{tab:model}. Note
that when fitting the data there is an additional parameter, $h$,
which is not considered as a model parameter but is included in $k$
when calculating AIC and BIC. The fit and information criteria
results are summarized in Table~\ref{tab:result}.

Next, we shall outline the basic equations describing the evolution
of the cosmic expansion in each of the competing dark energy models,
calculate the best-fit values of their parameters, and find their
corresponding $\chi^2_{min}$, $\Delta$AIC and $\Delta$BIC values.


\begin{table}
\caption{\label{tab:model} Summary of models} \footnotesize\rm
\begin{tabular*}{\textwidth}{@{}l*{15}{@{\extracolsep{0pt plus12pt}}l}}
\hline\hline
Model   &   Abbreviation\tablenotemark[1]   &  Model parameters\tablenotemark[2] ${\bm \theta}$   &   Number of model parameters ($k-1$)\\
\hline
Cosmological constant $\dotfill$   &   $\Lambda$   &   $\Omega_m$   &   1\\
 Constant $w$ $\dotfill$   &   $w$   &   $\Omega_m$, $w$   &   2\\
Chevallier-Polarski-Linder $\dotfill$  &  CPL  &  $\Omega_m$, $w_0$, $w_a$  &  3\\
Generalized Chaplygin gas  $\dotfill$  &  GCG  &  $A_s$, $\alpha$  &  2\\
Holographic dark energy$\dotfill$ & HDE & $\Omega_m$, $c$ & 2\\
Agegraphic dark energy$\dotfill$ & ADE & $n$ & 1 \\
Ricci dark energy $\dotfill$ & RDE & $\Omega_m$, $\alpha$ & 2\\
Dvali-Gabadadze-Porrati $\dotfill$ & DGP & $\Omega_m$ & 1\\
Phenomenological extension of DGP $\dotfill$ & $\alpha$DE & $\Omega_m$, $\alpha$ & 2\\
\hline \tablenotetext[1]{The abbreviations used in
Table~\ref{tab:result} and Fig.~\ref{AICBIC}.} \tablenotetext[2]{The
free parameters in each model. Note that the additional parameter
$h$ appearing in the data fits is not considered as a model
parameter but is included in $k$ when calculating AIC and BIC.}
\end{tabular*}
\end{table}

\begin{table}
\caption{\label{tab:result} Summary of the information criteria
results} \footnotesize\rm
\begin{tabular*}{\textwidth}{@{}l*{15}{@{\extracolsep{0pt plus12pt}}l}}
\hline\hline
Model         ~~~~~~~~&~~~  $\chi^2_{min}$~~&   ~~~~~~~~~~$\Delta$AIC   &    ~~~~~~~~$\Delta$BIC     ~~~~~~~~~~\\
\hline
$\Lambda$        ~~~~~~~~~&~~~    468.461    ~~& ~~~~~~~~~~0~~ &        ~~~~~~~~ 0           ~~~~~~~~~~\\
$w$       ~~~~~~~~~&~~~    468.327    ~~&~~~~~~~~~~1.866~~ &       ~~~~~~~~ 5.862         ~~~~~~~~~~\\
$\alpha$DE  ~~~~~~~~~&~~~    468.452    ~~&~~~~~~~~~~1.991   &      ~~~~~~~~ 5.987         ~~~~~~~~~~\\
GCG        ~~~~~~~~~&~~~    468.461    ~~&~~~~~~~~~~2   &      ~~~~~~~~ 5.996        ~~~~~~~~~~\\
CPL       ~~~~~~~~~&~~~    467.663    ~~&~~~~~~~~~~3.202   &       ~~~~~~~~ 11.195        ~~~~~~~~~~\\
HDE        ~~~~~~~~~&~~~    470.513    ~~&~~~~~~~~~~4.052   &      ~~~~~~~~ 8.048         ~~~~~~~~~~\\
RDE       ~~~~~~~~~&~~~   493.772     ~~&~~~~~~~~~~27.311   &      ~~~~~~~~ 31.308        ~~~~~~~~~~\\
ADE         ~~~~~~~~~&~~~    503.039    ~~&~~~~~~~~~~34.578   &      ~~~~~~~~ 34.578        ~~~~~~~~~~\\
DGP        ~~~~~~~~~&~~~    530.443    ~~&~~~~~~~~~~61.982~~ &      ~~~~~~~~ 61.982         ~~~~~~~~~~\\
\hline
\tablenotetext{Notes: The cosmological constant model is preferred
by both the AIC and the BIC. Thus, the $\Delta$AIC and $\Delta$BIC
values for all other models in the table are measured with respect
to these lowest values. The models are given in order of increasing
$\Delta$AIC.}
\end{tabular*}
\end{table}

\begin{figure}
\includegraphics[scale=0.3, angle=0]{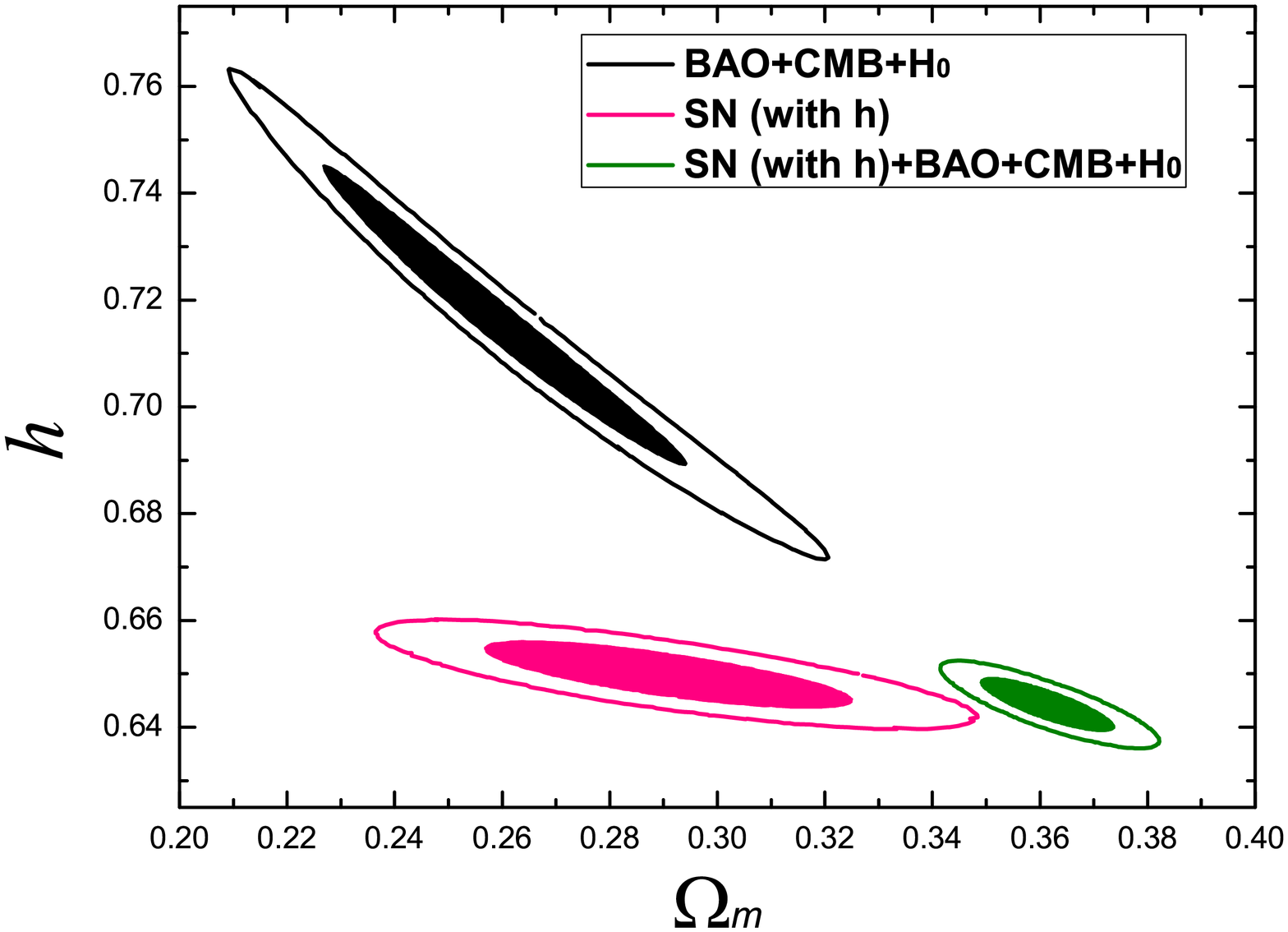}
\includegraphics[scale=0.3, angle=0]{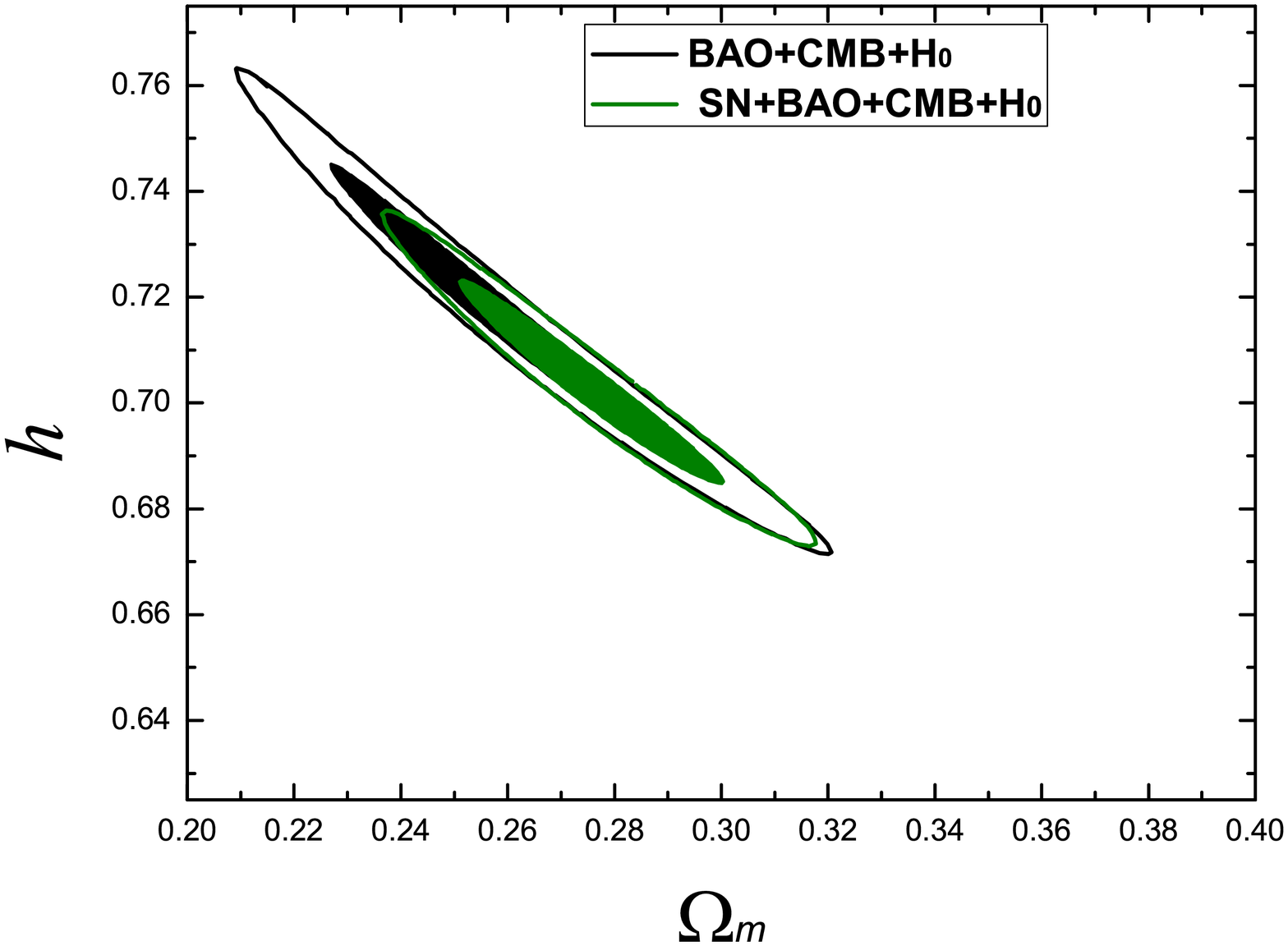}
\caption{The cosmological constant model: likelihood contours at
68.3\% and 95.4\% confidence levels in the $\Omega_m-h$ plane. The
left panel shows that when we use the SN Constitution data with $h$
not marginalized, there will be an inconsistency between the data of
SN and other cosmological probes. The right panel tells us that the
tension will disappear when we use the SN Constitution data with $h$
marginalized.} \label{figLCDM2}
\end{figure}

\subsection{Cosmological constant model}

The cosmological constant $\Lambda$ was first introduced by Einstein
\cite{Einstein1917} with a wrong motivation, but nowadays it has
become the most promising candidate for dark energy responsible for
the current cosmic acceleration. While it has been suffering from
the theoretical problems, it can explain the observations well. The
cosmological model containing a cosmological constant and cold dark
matter (CDM) component is usually called $\Lambda$CDM model in the
literature. The unique feature of the cosmological constant is that
its equation-of-state parameter $w$ has the value $-1$ at all times,
so in this model we have \be \label{LCDM}
E(z)=\sqrt{\Omega_{m}(1+z)^3+\Omega_{r}(1+z)^4+(1-\Omega_{m}-\Omega_{r})}.
\ee It is obvious that this model is a one-parameter model, with the
sole independent parameter $\Omega_m$.

The best-fit values of parameters (including the model parameter
$\Omega_m$ and the dimensionless Hubble parameter $h$) and the
corresponding $\chi^2_{min}$ are:
\begin{equation}
\Omega_m=0.275,\ \ \ \ h=0.704,\ \ \ \ \chi^2_{min}=468.461.
\end{equation}
This model has the lowest values of AIC and BIC in all the models
tested, so $\Delta$AIC and $\Delta$BIC are measured with respect to
this model; see Table~\ref{tab:result}.

To make a comparison, we refer to Ref. \cite{WMAP7} in which Komatsu
\textit{et al.} give the best-fit parameters: $\Omega_{m}=0.273$ and
$h=0.702$, for the flat $\Lambda$CDM model from WMAP 7-year data
combined with BAO and $H_0$ data. We find that our results with SN
Constitution data are consistent with this result. We also stress
that in the joint data analysis we have used the chi square of SN
data $\tilde{\chi}^2_{SN}$ that is $h$-free, instead of
$\chi^2_{SN}$ that is $h$-relevant. For making a clear comparison,
we also perform a joint analysis with the $h$-relevant SN chi square
$\chi^2_{SN}$, and in this way we find the best-fit parameters:
$\Omega_m=0.361$ and $h=0.644$, with $\chi^2_{min}=495.55$, in
accordance with the results of Ref.~\cite{Gong:2009ye}. Such a big
$\chi^2_{min}$ implies that this way does not seem to be correct.
Figure \ref{figLCDM2} shows the probability contours in the
$\Omega_m-h$ plane for the flat $\Lambda$CDM model. The left panel
tells us that if we use the $h$-relevant $\chi^2_{SN}$, a great
tension will be brought between the SN limit and the BAO+CMB limit,
and the right panel shows that the tension will disappear when
considering the $h$-free $\tilde{\chi}^2_{SN}$. Actually, in
Ref.~\cite{Wang:2007mza}, Wang and Mukherjee have argued that
because of calibration uncertainties, SN data need to be
marginalized over $h$ if SN data are combined with data that are
sensitive to the value of $h$. Our results further confirm this
opinion.

\subsection{Dark energy models with equation of state parameterized}

For this class, we consider two models: the constant $w$
parametrization and the Chevallier-Polarski-Linder (CPL)
parametrization.

\subsubsection{Constant $w$ parametrization}

In this case, the equation-of-state parameter of dark energy is
assumed to be a constant, so in a flat universe we have \be
\label{XCDM}
E(z)=\sqrt{\Omega_{m}(1+z)^3+\Omega_{r}(1+z)^4+(1-\Omega_{m}-\Omega_{r})(1+z)^{3(1+w)}}.
\ee This is a two-parameter model with the model parameters
$\Omega_m$ and $w$. 

\begin{figure}
\includegraphics[scale=0.3, angle=0]{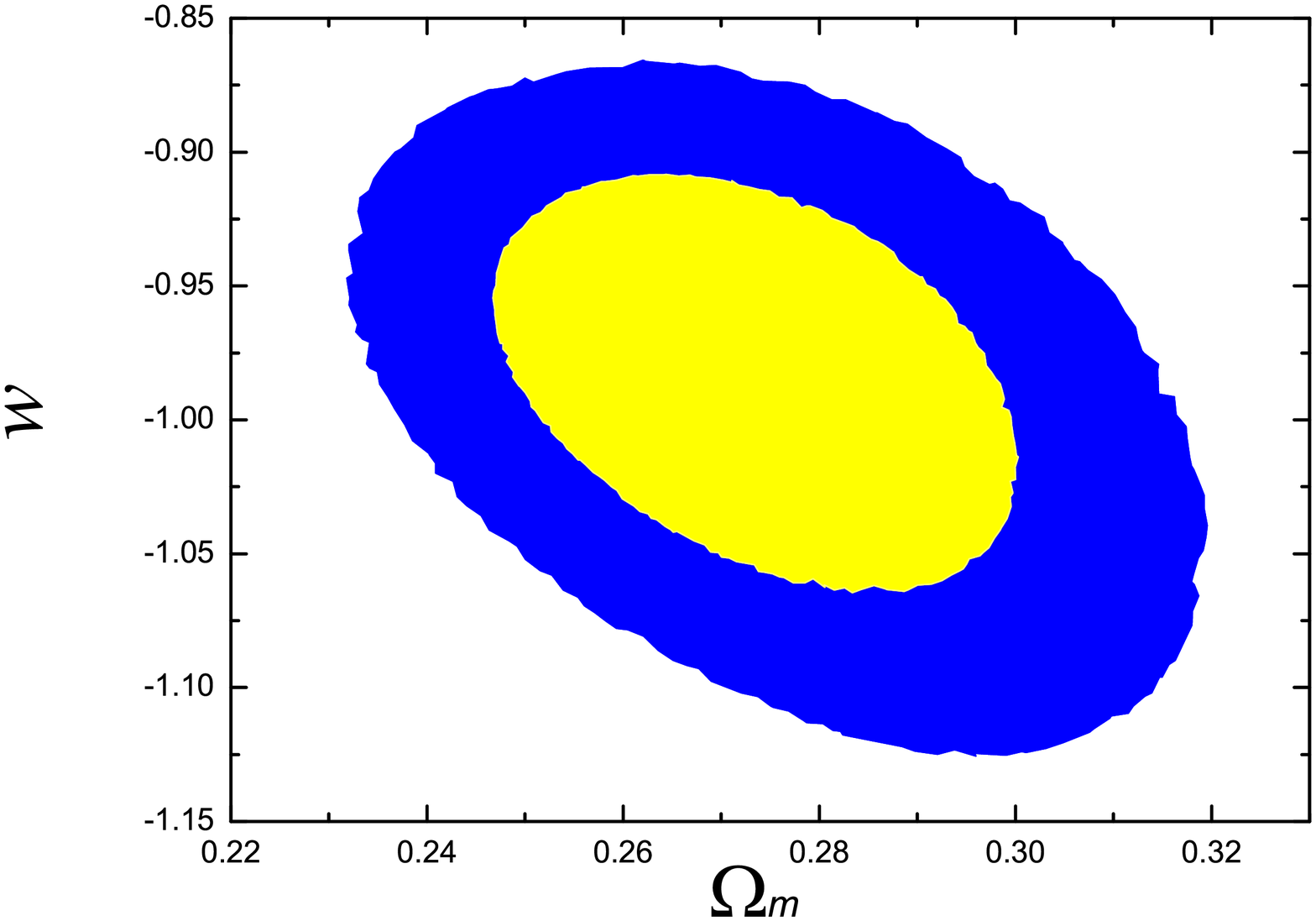}
\includegraphics[scale=0.3, angle=0]{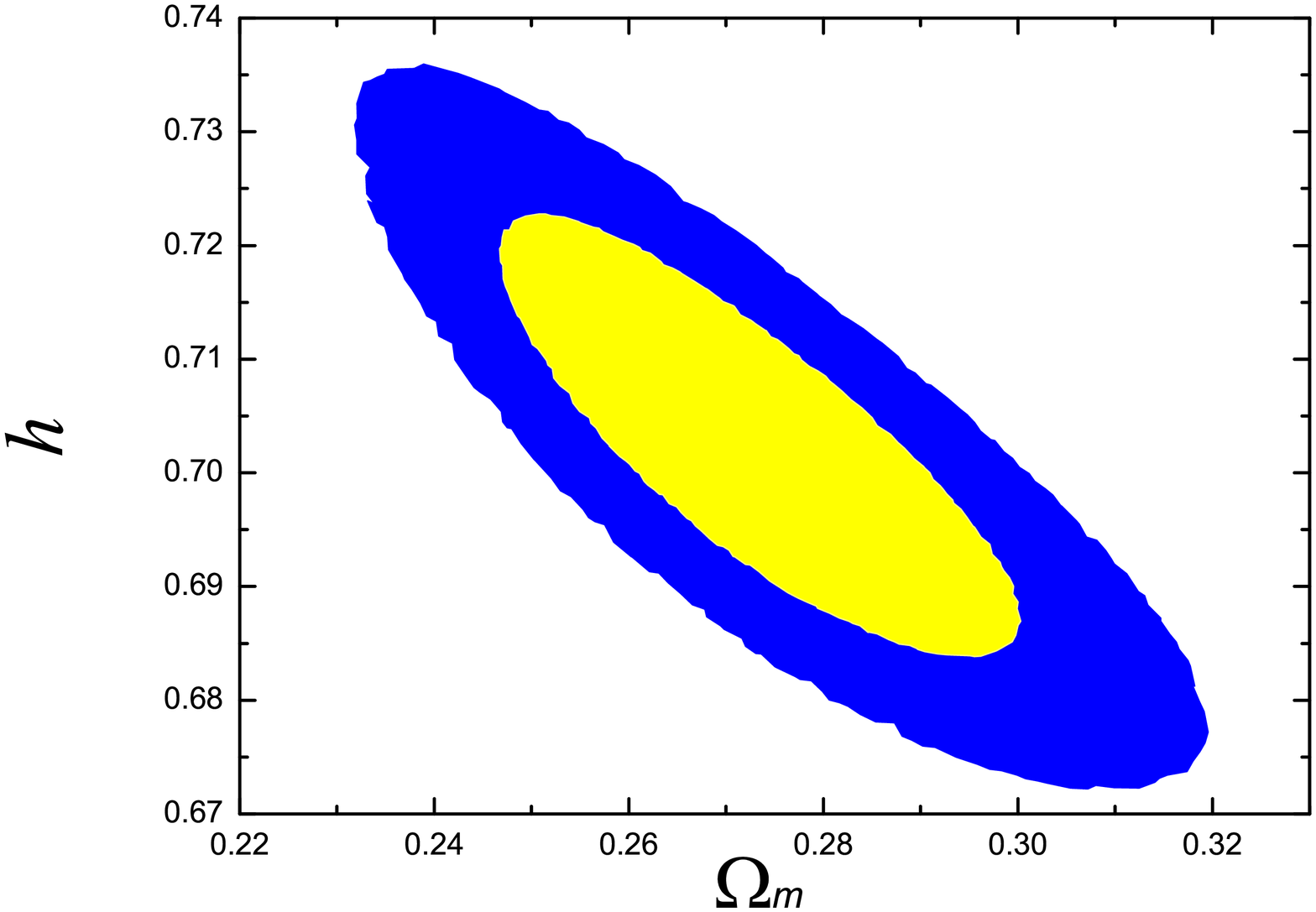}
\caption{The constant $w$ model: likelihood contours at $68.3\%$ and
$95.4\%$ confidence levels in the $\Omega_m-w$ and $\Omega_m-h$
planes.} \label{figXCDM}
\end{figure}

The best-fit parameters and the corresponding $\chi^2_{min}$ are:
\begin{equation}
\Omega_m=0.272,\ \ \ \ w=-0.981,\ \ \ \ h=0.703,\ \ \ \
\chi^2_{min}=468.327.
\end{equation}
We plot the likelihood contours for this model in the $\Omega_m-w$
and $\Omega_m-h$ planes in Fig.~\ref{figXCDM}. From this figure, we
see that when the equation of state does not depend on redshifts,
dark energy is consistent with a cosmological constant within
1$\sigma$ range. Comparing with the cosmological constant model,
this model gives a lower $\chi_{min}^2$, but due to one extra
parameter it has, it is punished by the information criteria:
$\Delta{\rm AIC}=1.866$ and $\Delta{\rm BIC}=5.862$.

\subsubsection{Chevallier-Polarski-Linder parametrization}

Now, we consider the commonly used CPL model \cite{CPL}, in which
the equation of state of dark energy is parameterized as
\begin{equation}
\label{cpl1} w(z)=w_0+w_a\frac{z}{1+z},
\end{equation}
where $w_0$ and $w_a$ are constants. The corresponding $E(z)$ can be
expressed as
\begin{equation} \label{cpl2}
E(z)=\sqrt{\Omega_{m}(1+z)^3+\Omega_{r}(1+z)^4+(1-\Omega_{m}-\Omega_{r})(1+z)^{3(1+w_0+w_a)}\exp\left(-\frac{3w_a
z}{1+z}\right)}.
\end{equation}
There are three independent model parameters in this model:
${\bm\theta}=\{\Omega_m,~w_0,~w_a\}$. 

\begin{figure}
\includegraphics[scale=0.3, angle=0]{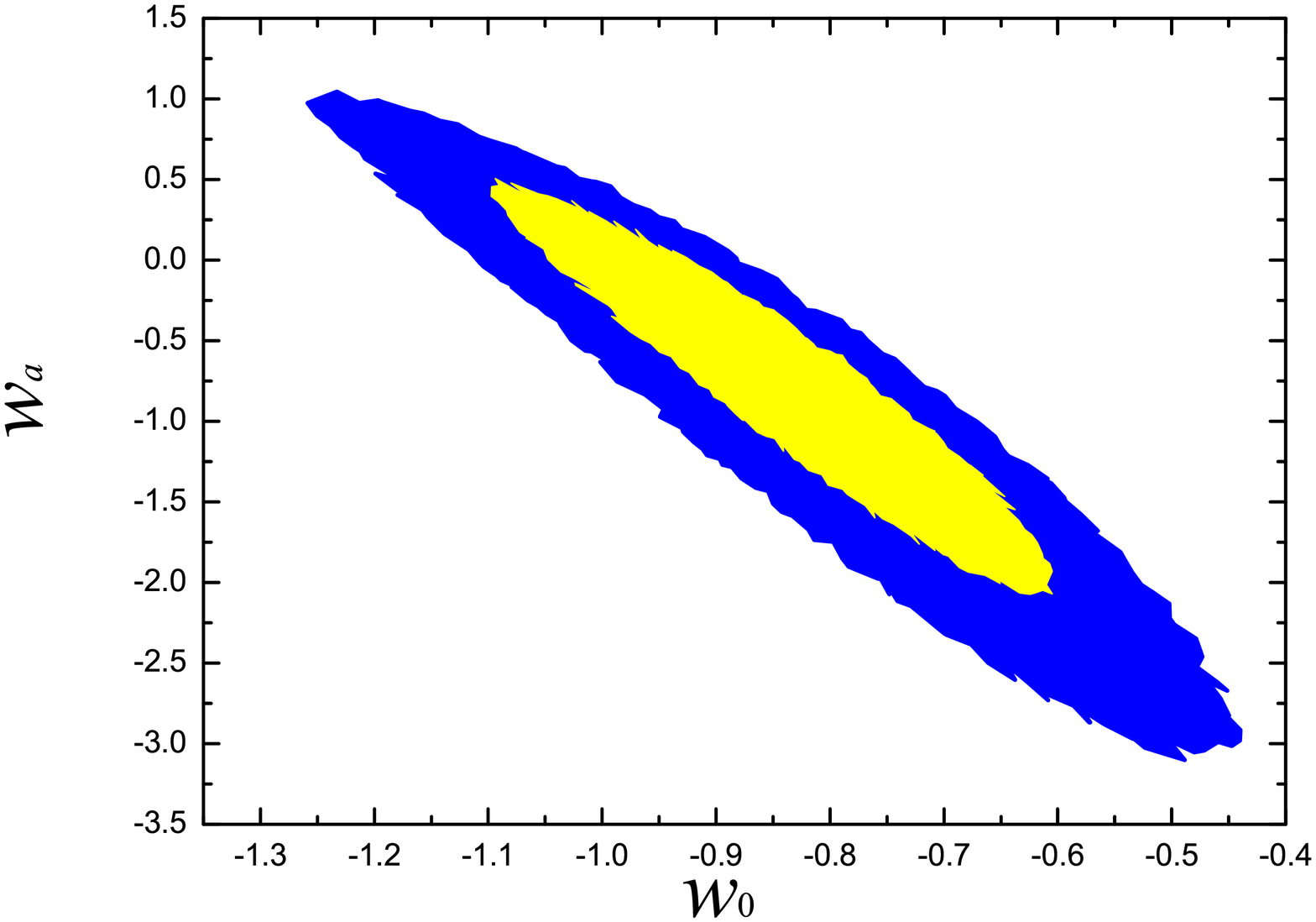}
\includegraphics[scale=0.3, angle=0]{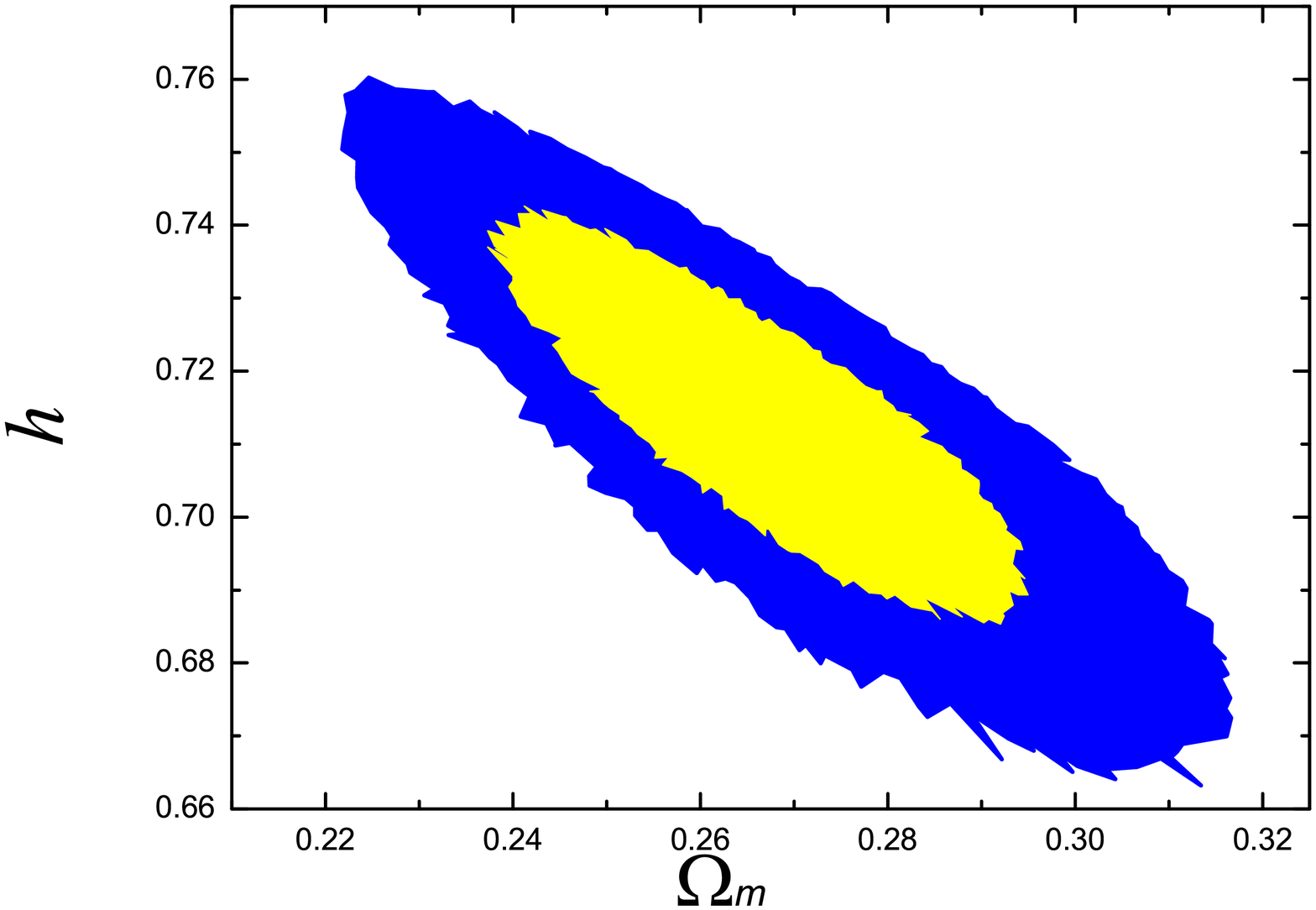}
\caption{The Chevallier-Polarski-Linder (CPL) model: likelihood
contours at 68.3\% and 95.4\% confidence levels in the $w_0-w_a$ and
$\Omega_m-h$ planes.} \label{figCPL}
\end{figure}

According to the joint data analysis, we find the best-fit
parameters and the corresponding $\chi^2_{min}$:
\begin{equation}
\Omega_m=0.265,\ \ \ \ \  w_0 =-0.847,\ \ \ \ \  w_a =-0.691,\ \ \ \
\ h=0.716,\ \ \ \ \ \ \chi^2_{min}=467.663.
\end{equation}
We plot the likelihood contours for the CPL model in the $w_0-w_a$
and $\Omega_m-h$ planes in Fig.~\ref{figCPL}. We note that the
best-fit parameters of the $\Lambda$CDM model ($\Omega_m=0.275$,
$w_0=-1$ and $w_a=0$) still lie in the 1$\sigma$ regions of the CPL
model, indicating that the $\Lambda$CDM model is fairly consistent
with the current observational data. Since the CPL model has three
free model parameters, it should have made considerable improvement
in the fit, however, it gives a nearly equal $\chi^2_{min}$
contrasting to the constant $w$ model (only smaller by 0.664). The
differences in the information criteria with respect to the
$\Lambda$CDM model are: $\Delta{\rm AIC}=3.202$ and $\Delta{\rm
BIC}=11.195$. The information criteria result of CPL is worse than
the $w$ model (especially its $\Delta{\rm BIC}$ value is very
large). This implies that the CPL model is too complex to be
necessary in explaining the current data, comparing with the simpler
models such as the $\Lambda$ model and the constant $w$ model.

\subsection{Chaplygin gas models}

Chaplygin gas models describe a background fluid with
$p\propto\rho^{-\alpha}$ that is commonly viewed as arising from the
d-brane theory. Moreover, these models may be able to unify dark
energy and dark matter. We should have considered both the original
($\alpha=1$) and the generalized Chaplygin gas models, however, the
original Chaplygin gas model \cite{cg} has been proven to be
inconsistent with the observational data \cite{Davis:2007na}, we
thus only consider the generalized Chaplygin gas (GCG) model
\cite{gcg} in this paper.

The GCG has an exotic equation of state: \be \label{gcg}
p_{gcg}=-\frac{A}{\rho_{gcg}^{\alpha}}, \ee where $A$ is a positive
constant. This leads to the energy density of the GCG:
\begin{equation}
\rho_{gcg}(a)=\rho_{gcg}(0)\left(A_s+{1-A_s\over
a^{3(1+\alpha)}}\right)^{1\over 1+\alpha},
\end{equation}
where $A_s\equiv A/\rho_{gcg}^{1+\alpha}(0)$. When $A_s=0$, the GCG
behaves like a dust matter; when $A_s=1$, the GCG behaves like a
cosmological constant. So, the GCG model is considered as a
unification scheme of the cosmological constant and the CDM. In a
flat universe, we have \be \label{GCG}
E(z)=\sqrt{\Omega_{b}(1+z)^3+\Omega_{r}(1+z)^4+(1-\Omega_{b}-\Omega_{r})\left(A_s+(1-A_s)(1+z)^{3(1+\alpha)}\right)^{1/1+\alpha}}.
\ee This model has two independent model parameters:
${\bm\theta}=\{A_s,~\alpha\}$. The cosmological constant model is
recovered for $\alpha=0$ and
$\Omega_m=1-\Omega_{r}-A_s(1-\Omega_{r}-\Omega_{b})$.

\begin{figure}
\includegraphics[scale=0.3, angle=0]{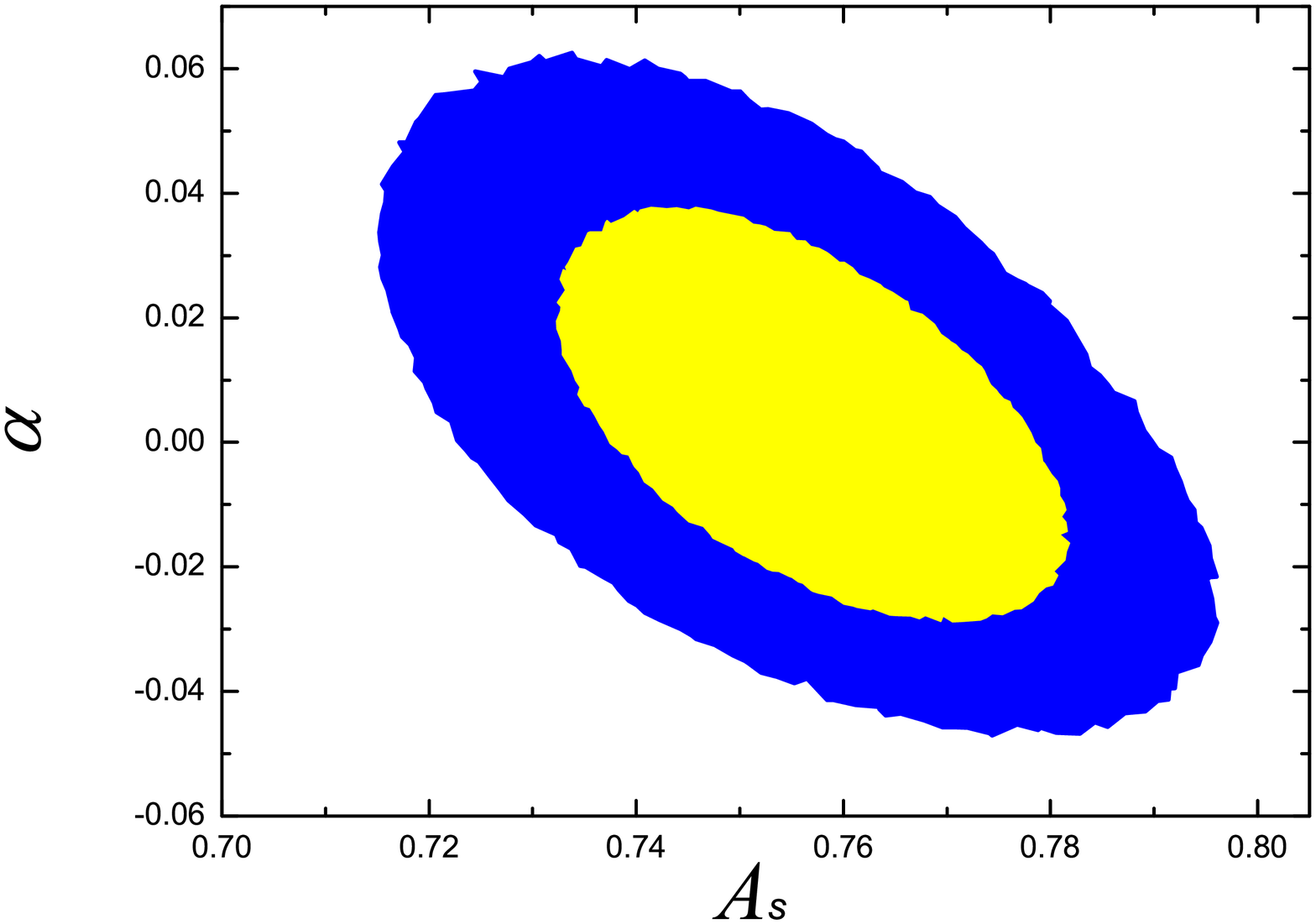}
\includegraphics[scale=0.3, angle=0]{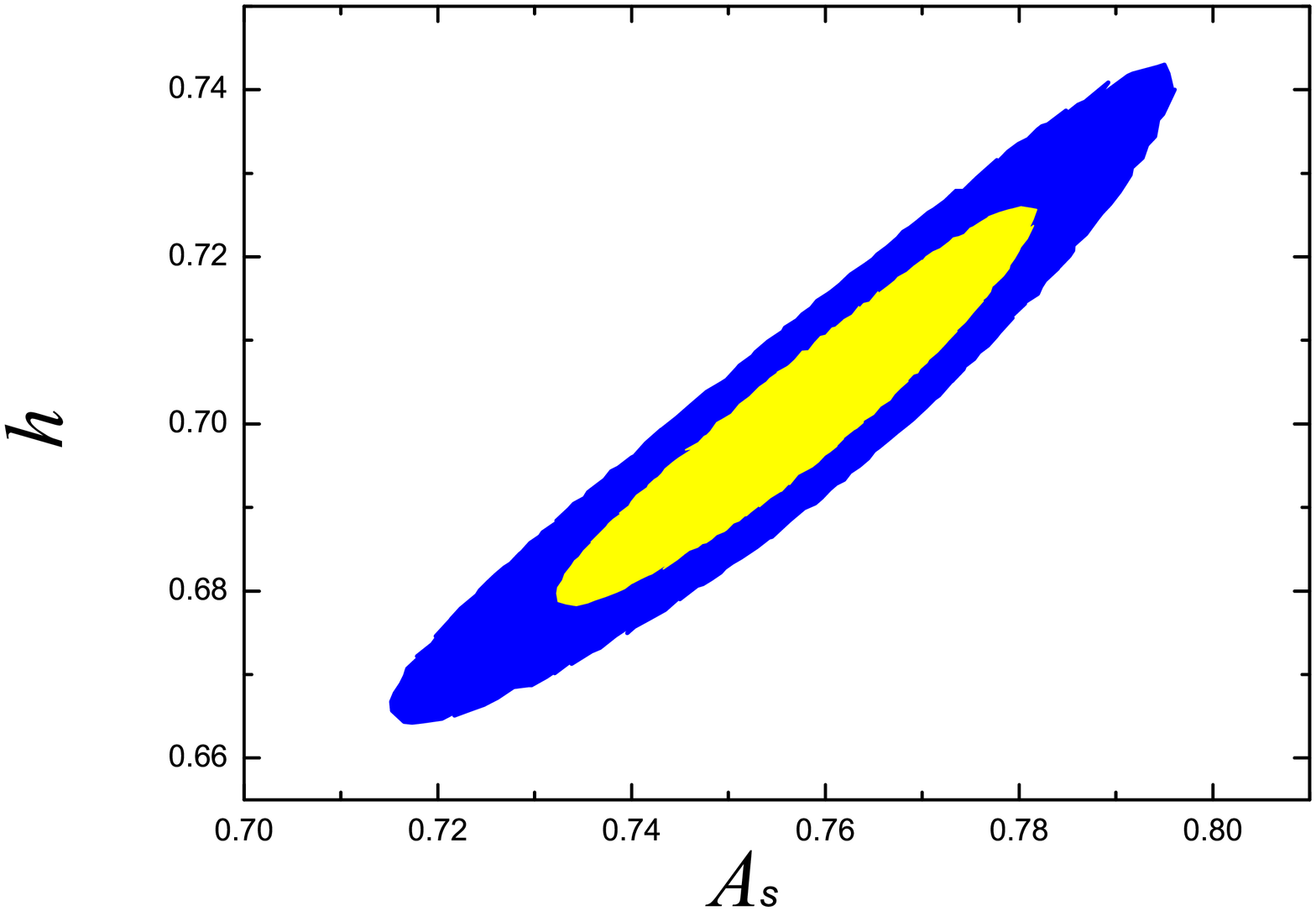}
\caption{The generalized Chaplygin gas model: likelihood contours at
68.3\% and 95.4\% confidence levels in the $A_s-\alpha$ and $A_s-h$
planes.} \label{figCG}
\end{figure}

The best-fit parameters and the corresponding $\chi^2_{min}$ are:
\begin{equation}
A_s=0.758,\ \ \ \ \ \ \alpha=0.003,\ \ \ \ \ h=0.701,\ \ \ \ \
\chi^2_{min}=468.461.
\end{equation}
We find that for the GCG model the $\chi^2_{min}$ value is the same
as that of the $\Lambda$CDM model, which is an amazing coincidence.
The best-fit value of $\alpha$ is so close to zero, implying that
the $\Lambda$CDM limit of this model is favored. We plot the
likelihood contours for the GCG model in the $A_s-\alpha$ and
$A_s-h$ planes in Fig.~\ref{figCG}. As a two-parameter model, the
GCG performs well under the information criteria tests: $\Delta{\rm
AIC}=2$ and $\Delta{\rm BIC}=5.996$.

\subsection{Holographic dark energy models}

Holographic dark energy models arise from the holographic principle
of quantum gravity. The holographic principle determines the range
of validity for a local effective quantum field theory to be an
accurate description of the world involving dark energy, by imposing
a relationship between the ultraviolet (UV) and infrared (IR)
cutoffs \cite{Cohen99}. As a consequence, the vacuum energy becomes
dynamical, and its density $\rho_{de}$ is inversely proportional to
the square of the IR cutoff length scale $L$ that is believed to be
some horizon size of the universe, namely, $\rho_{de}\propto
L^{-2}$. In this subsection, we consider three holographic dark
energy models: the original holographic dark energy (HDE) model
\cite{hde}, the agegraphic dark energy (ADE) model \cite{ade}, and
the holographic Ricci dark energy (RDE) model \cite{rde}. We note
here that, different from the previous several models, the
holographic dark energy models do not involve the cosmological
constant model as a subclass.

\subsubsection{Holographic dark energy model}

The HDE model chooses the future event horizon size as its IR cutoff
scale, so the energy density of HDE reads
\begin{equation}
\rho_{de}=3c^2M^2_{Pl}R_{eh}^{-2},\label{hde}
\end{equation}
where $c$ is a constant, and $R_{eh}$ is the size of the future
event horizon of the universe, defined as
 \be \label{Reh}
R_{eh} =
a\int_{t}^{\infty}\frac{dt'}{a}=a\int_{a}^{\infty}\frac{da'}{Ha'^2}.
\ee

In this case, $E(z)$ is given by Eq.~(\ref{Ez2}), where the function
$\Omega_{de}(z)$ is determined by a differential equation:
\begin{equation}
\label{HDE} \Omega'_{de}(z)=-\frac{2\Omega_{de}(z)}{1+z} \left(
\epsilon(z)-1+\frac{\sqrt{\Omega_{de}(z)}}{c} \right),
\end{equation}
where a prime denotes $d/dz$, and
\begin{equation}
\label{epsilon} \epsilon(z)={3\over 2}\left[{1+{4\over 3}(1+z)\gamma
\over
1+(1+z)\gamma}(1-\Omega_{de}(z))+(1+w_{de}(z))\Omega_{de}(z)\right],
\end{equation}
with
\begin{equation}
\gamma=\Omega_{r}/\Omega_{m},\ \
w_{de}(z)=-\frac{1}{3}-\frac{2}{3}\sqrt{\Omega_{de}(z)}.
\end{equation}
The HDE model contains two independent model parameters:
${\bm\theta}=\{\Omega_m,~c\}$. Solving Eq. (\ref{HDE}) numerically
and substituting the resultant $\Omega_{de}(z)$ into Eq.
(\ref{Ez2}), the corresponding $E(z)$ can be obtained.

\begin{figure}
\includegraphics[scale=0.3, angle=0]{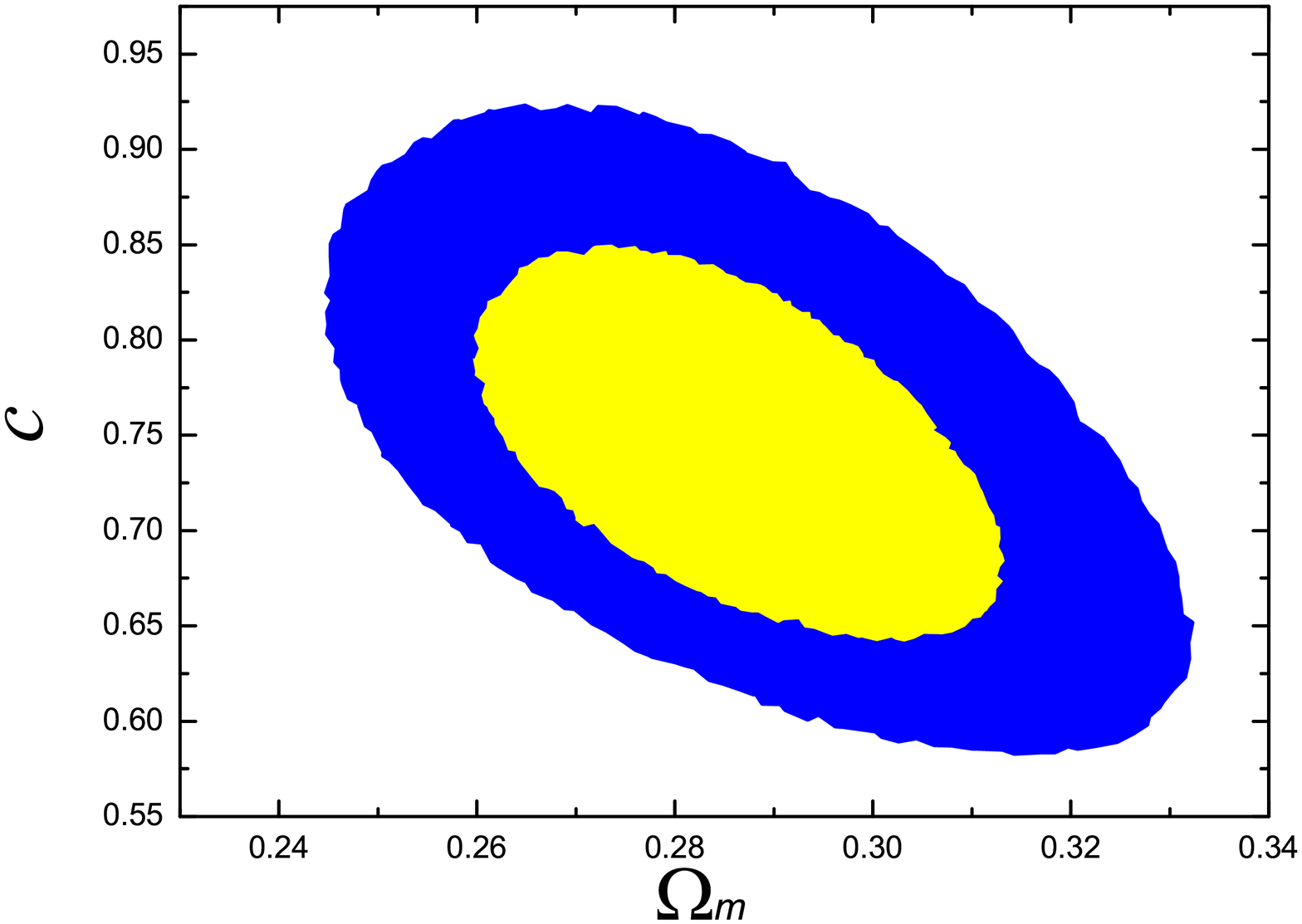}
\includegraphics[scale=0.3, angle=0]{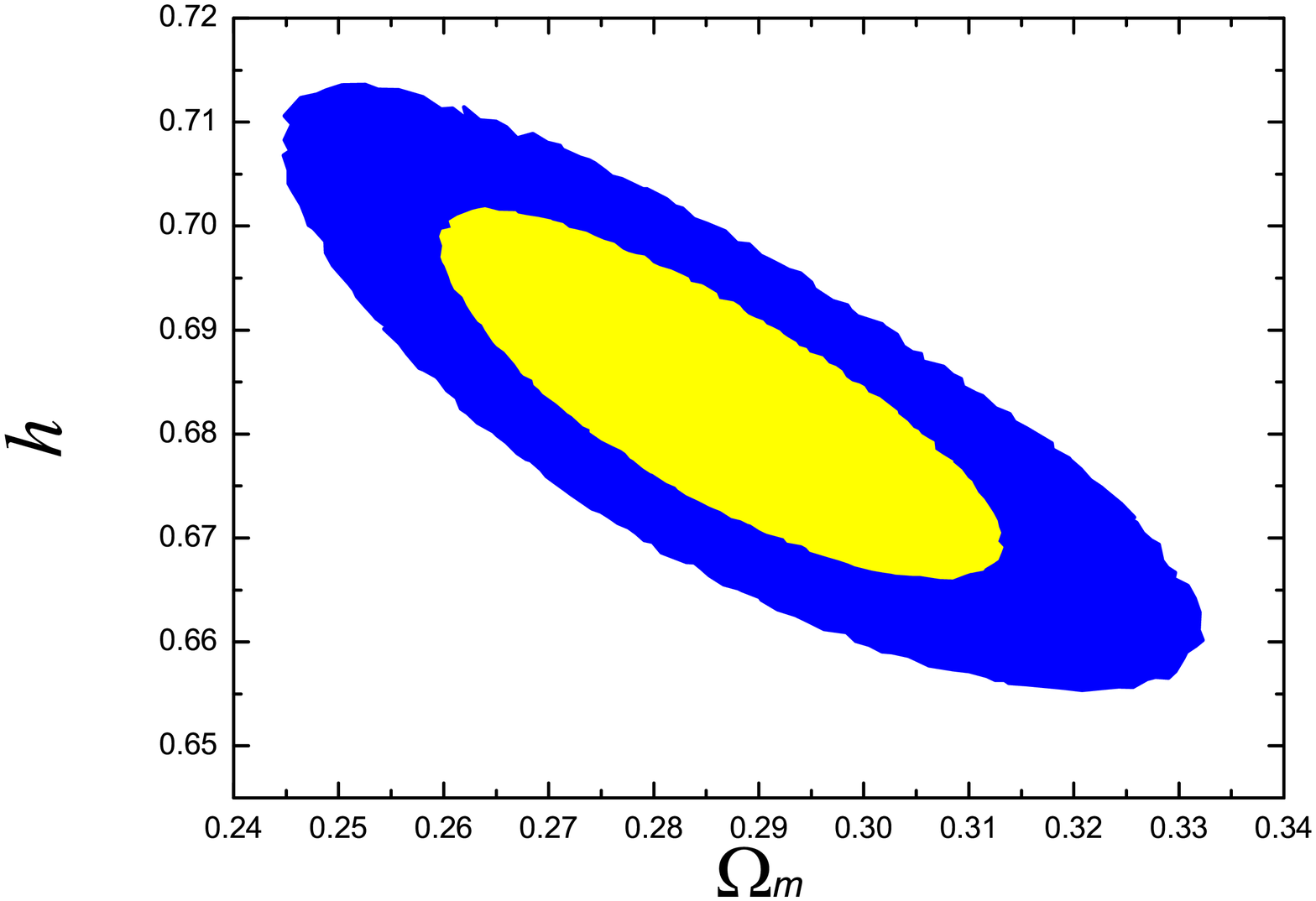}
\caption{The holographic dark energy model: likelihood contours at
68.3\% and 95.4\% confidence levels in the $\Omega_m-c$ and
$\Omega_m-h$ planes.} \label{figHDE}
\end{figure}

For this model, we obtain the best-fit parameters and the
corresponding $\chi^2_{min}$:
\begin{equation}
\Omega_m=0.285,\ \ \ \ \ c=0.742,\ \ \ \ \ h=0.684, \ \ \ \ \ \
 \chi^2_{min}=470.513.
\end{equation}
Our results are generally consistent with those derived in previous
works \cite{holomodels09,holofitzx}. We plot the likelihood contours
for the HDE model in the $\Omega_m-c$ and $\Omega_m-h$ planes in
Fig.~\ref{figHDE}. The HDE model performs fine under the information
criteria tests, with the results $\Delta{\rm AIC}=4.052$ and
$\Delta{\rm BIC}=8.048$. It should be stressed that the HDE model
does not contain the $\Lambda$CDM model as a sub-model, whereas
other two-parameter models which perform better than the HDE model,
such as the $\alpha$DE, constant $w$, and GCG models, all nest the
cosmological constant $\Lambda$ and tend to collapse to the
$\Lambda$CDM model once being up against the current observational
data.

\subsubsection{Agegraphic dark energy model}

The ADE model discussed in this paper is actually the new version of
the ADE model \cite{ade} (sometimes called the new ADE model in the
literature) which chooses the conformal age of the universe
\begin{equation}
\eta=\int_0^t \frac{dt'}{a}=\int_0^a \frac{da'}{Ha'^{2}}
\end{equation}
as the IR cutoff, so the energy density of ADE is
\begin{equation}
\label{ade} \rho_{de}=3n^{2}M_{Pl}^{2}\eta^{-2},
\end{equation}
where $n$ is a constant which plays the same role as $c$ in the HDE
model.

As the same as the HDE model, $E(z)$ is also given by
Eq.~(\ref{Ez2}), where the function $\Omega_{de}(z)$ is governed by
the differential equation:
\begin{equation}
\label{ADE} \Omega'_{de}(z)=-\frac{2\Omega_{de}(z)}{1+z} \left(
\epsilon(z)-\frac{(1+z)\sqrt{\Omega_{de}(z)}}{n} \right),
\end{equation}
where the form of $\epsilon(z)$ is the same as Eq.~(\ref{epsilon}),
in which
\begin{equation}
w_{de}(z)=-1+\frac{2(1+z)\sqrt{\Omega_{de}(z)}}{3n}.
\end{equation}
Following Ref.~\cite{Wei:2007xu}, we choose the initial condition,
$\Omega_{de}(z_{ini})=n^2(1+z_{ini})^{-2}/4$, at $z_{ini}=2000$, and
then Eq.~(\ref{ADE}) can be numerically solved. Substituting the
resultant $\Omega_{de}(z)$ into Eq.~(\ref{Ez2}), the function $E(z)$
can be obtained. Note that in this model once $n$ is given, by
solving Eq.~(\ref{ADE}), $\Omega_{m}=1-\Omega_{de}(0)-\Omega_{r}$
can be derived accordingly. So, actually, the ADE model is a
one-parameter model; the sole model parameter is $n$.

\begin{figure}
\includegraphics[scale=0.3, angle=0]{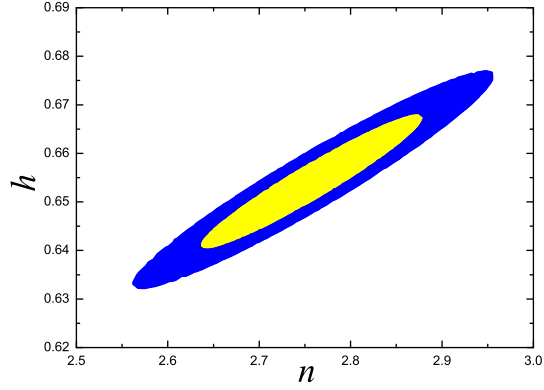}
\caption{The agegraphic dark energy model: likelihood contours at
68.3\% and 95.4\% confidence levels in the $n-h$ plane.}
\label{figADE}
\end{figure}

For this model, we get the best-fit parameters and the corresponding
$\chi^2_{min}$:
\begin{equation}
n =2.755,\ \ \ \ \  h =0.654,\ \ \ \ \  \chi^2_{min}=503.039.
\end{equation}
This leads to $\Omega_m=0.287$. We plot the likelihood contours for
the ADE model in the $n-h$ plane in Fig.~\ref{figADE}. As a
single-parameter model, the ADE performs much worse than the
$\Lambda$CDM model: its $\chi^2_{min}$ is greater than that of the
$\Lambda$CDM model by about 30, and its $\Delta{\rm AIC}=\Delta{\rm
BIC}=34.578$.

\subsubsection{Ricci dark energy model}

In the RDE model, the IR cutoff length scale is given by the average
radius of the Ricci scalar curvature $|{\cal R}|^{-1/2}$, so in this
case we have $\rho_{de}\propto {\cal R}$. In a flat universe, the
Ricci scalar is ${\cal R}=-6(\dot{H}+2H^2)$, and as suggested in
Ref.~\cite{rde}, the energy density of RDE reads
\begin{equation}
\rho_{de}=3\alpha M_{Pl}^2(\dot{H}+2H^2),\label{rde}
\end{equation}
where $\alpha$ is a positive constant. From the Friedmann equation,
we derive
\begin{equation}
E^2=\Omega_{m0}e^{-3x}+\Omega_{r0}e^{-4x}+\alpha \left({1\over
2}{dE^2\over dx}+2E^2\right),
\end{equation}
where $x=\ln a$. Solving this differential equation, we get the
following form: \be \label{RDE} E(z)=\sqrt{\frac{
2\Omega_{m}}{2-\alpha}(1+z)^{3}+\Omega_{r}(1+z)^4+(1-\Omega_{r}-{2\Omega_{m}\over
2 -\alpha})(1+z)^{(4-{2\over\alpha})}}. \ee This is a two-parameter
model, and its free model parameters are:
${\bm\theta}=\{\Omega_m,~\alpha\}$.

\begin{figure}
\includegraphics[scale=0.3, angle=0]{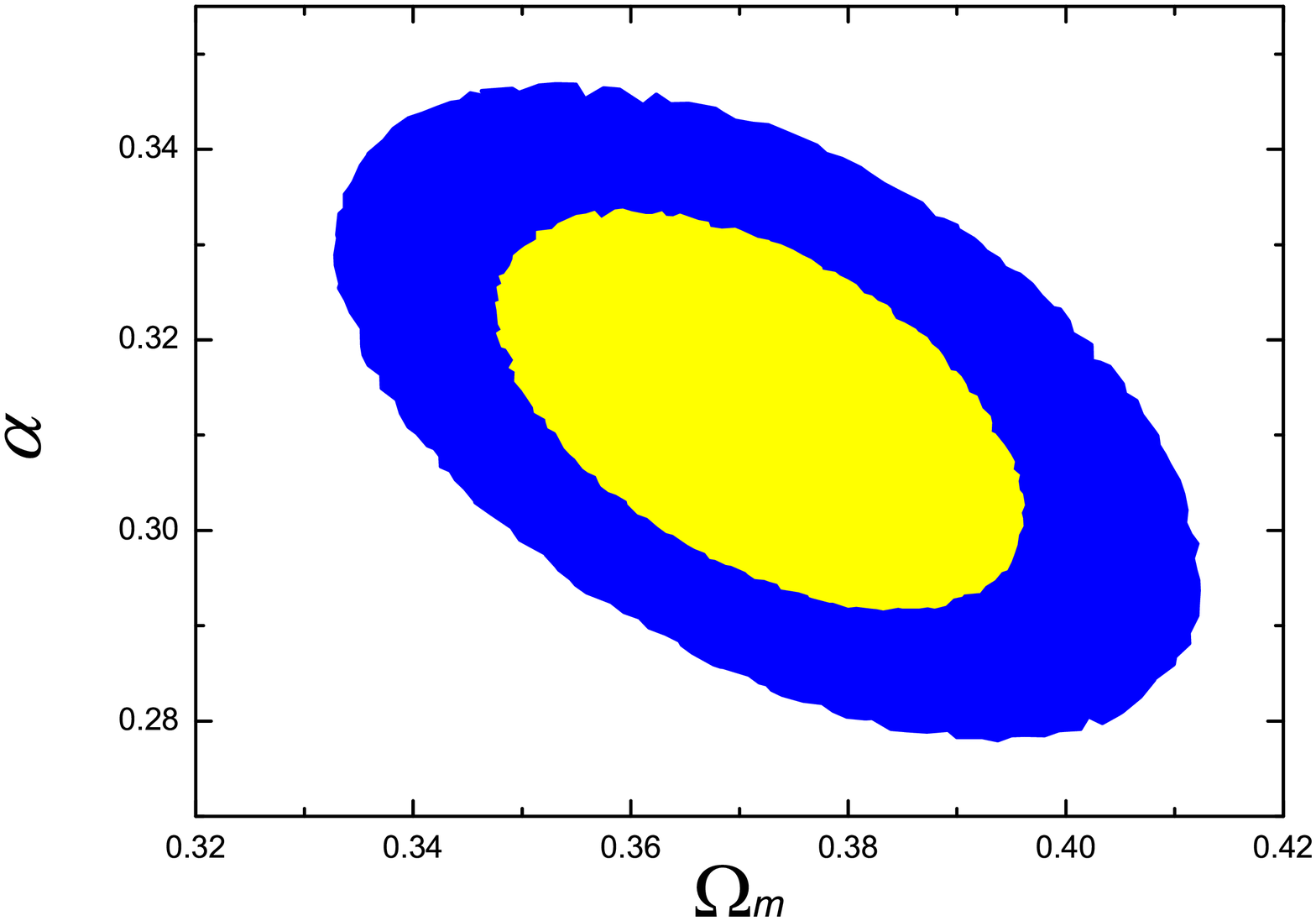}
\includegraphics[scale=0.3, angle=0]{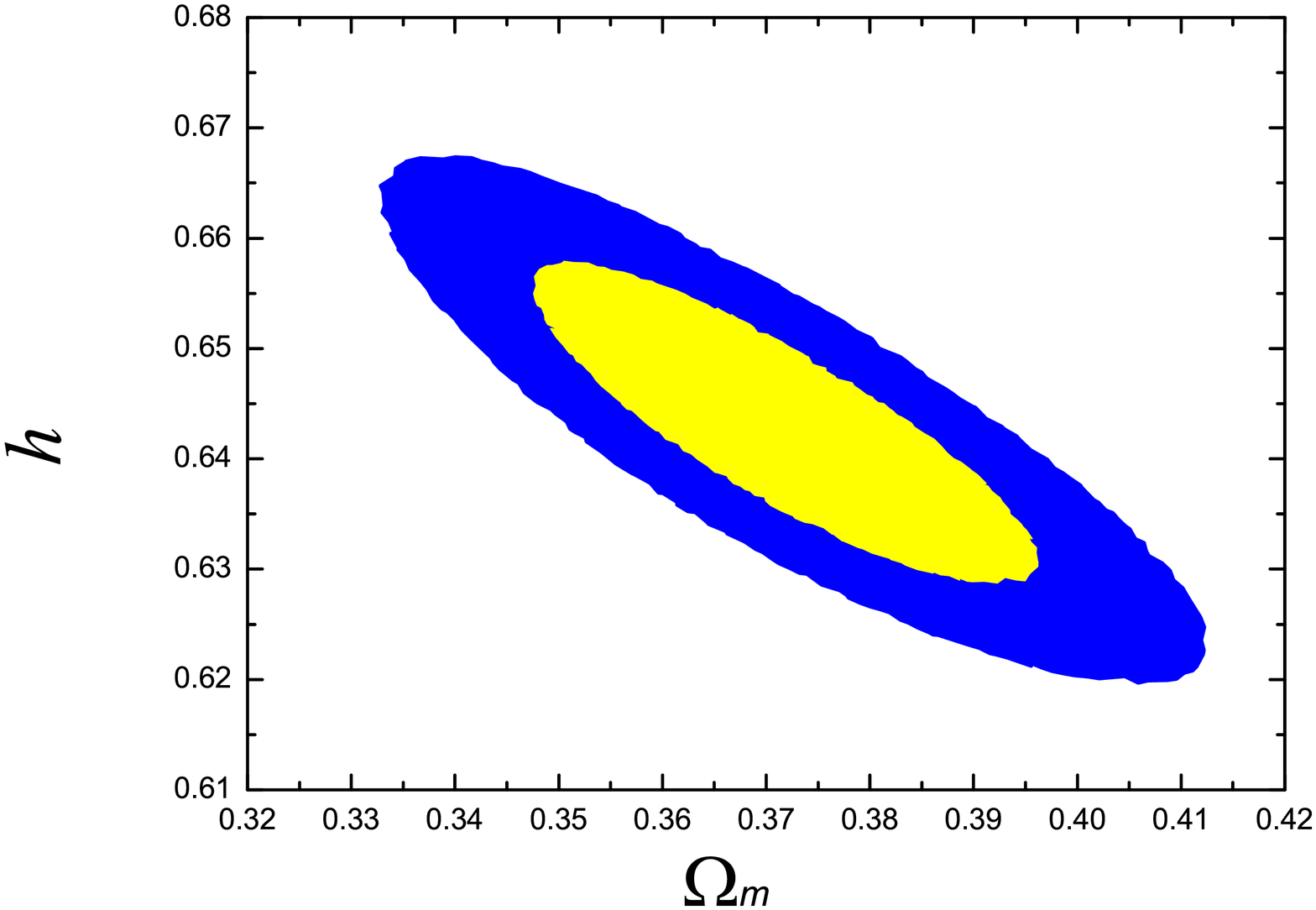}
\caption{The Ricci dark energy model: likelihood contours at 68.3\%
and 95.4\% confidence levels in the $\Omega_m-\alpha$ and
$\Omega_m-h$ planes.} \label{figRDE}
\end{figure}

For the RDE model, the best-fit parameters and the corresponding
$\chi^2_{min}$ are:
\begin{equation}
\Omega_m=0.371, \ \ \ \ \ \alpha =0.313,\ \ \ \ \ h =0.643, \ \ \ \
\ \chi^2_{min}=493.772.
\end{equation} We plot the likelihood contours for the RDE model in
the $\Omega_m-\alpha$ and $\Omega_m-h$ planes in Fig.~\ref{figRDE}.
Like the ADE model, RDE also performs very bad: $\Delta{\rm
AIC}=27.311$ and $\Delta{\rm BIC}=31.308$. This conclusion is
consistent with the previous work \cite{holomodels09,WangS}.

\subsection{Dvali-Gabadadze-Porrati brane world and related models}

The DGP brane world model is a well-known example of the
modification of general relativity for explaining the acceleration
of the universe. In this subsection, we consider two models: the DGP
model \cite{DGP} and its phenomenological extension (namely, the
$\alpha$ dark energy model) \cite{alphaDE}.

\subsubsection{Dvali-Gabadadze-Porrati model}

The DGP model arises from the brane world theory in which gravity
leaks out into the bulk at large scales, resulting in the
possibility of an accelerated expansion of the universe. In this
model, the Friedmann equation is modified as
\begin{equation}
3M_{Pl}^2\left(H^2-{H\over r_c}\right)=\rho_m(1+z)^3+\rho_r(1+z)^4,
\end{equation}
where $r_c=(H_0(1-\Omega_m-\Omega_r))^{-1}$ is the crossover scale.
In this model, $E(z)$ is given by \be \label{DGP}
E(z)=\sqrt{\Omega_{m}(1+z)^3+\Omega_{r}(1+z)^4+\Omega_{r_c}}+\sqrt{\Omega_{r_c}},
\ee where $\Omega_{rc}=1/(4r_c^2H_0^2)$ is a constant. The flat DGP
model only contains one free model parameter, $\Omega_m$.

\begin{figure}
\includegraphics[scale=0.3, angle=0]{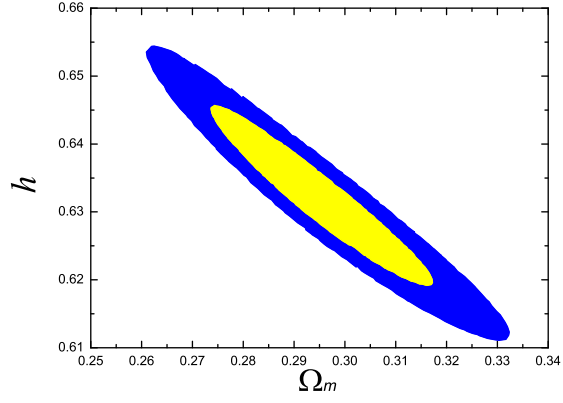}
\caption{The Dvali-Gabadadze-Porrati (DGP) model: likelihood
contours at 68.3\% and 95.4\% confidence levels in the $\Omega_m-h$
plane.} \label{figDGP}
\end{figure}

For the DGP model, the best-fit parameters and the corresponding
$\chi^2_{min}$ are:
\begin{equation}
\Omega_{m}=0.295,\ \ \ \ h=0.632,\ \ \ \ \chi^2_{min}=530.443.
\end{equation}
We plot the likelihood contours for the DGP model in the
$\Omega_m-h$ plane in Fig.~\ref{figDGP}. We see that the DGP model,
as a single-parameter model, is even worse than the ADE model under
the observational tests. Its $\chi^2_{min}$ is greater than that of
the ADE model by about 30, and it yields $\Delta{\rm AIC}=\Delta{\rm
BIC}=61.982$, also much larger than all other models we considered.
So, the fitting result shows that the DGP model seems to be
inconsistent with the current observational data (see also
Ref.~\cite{Davis:2007na}).

What should be mentioned is that the DGP model could perform much
better when considering the systematic errors of the SN Ia data. For
example, using the MLCS2k2 light-curve fitter for the SNe Ia data,
the authors of Ref. \cite{SDSS2} found that the DGP model performs
better than the standard $\Lambda$CDM model. Currently the SNe Ia
measurement errors are being dominated by systematic rather than
statistical uncertainties, and for the sake of simplicity we would
not discuss this problem in this paper. See Refs.~
\cite{SN09,SDSS2,Kessler,YunGuiSN} for detailed discussions of this
issue.

\subsubsection{Phenomenological extension of DGP: $\alpha$ dark energy model}

Inspired by the DGP brane world model in which the Friedmann
equation is modified, Dvali and Turner \cite{alphaDE} proposed a
phenomenological dark energy model, sometimes called the $\alpha$
dark energy model, which interpolates between the pure $\Lambda$CDM
model and the DGP model with an additional parameter $\alpha$. In
this model, the Friedmann equation is modified as
\begin{equation}
3M_{Pl}^2\left(H^2-{H^\alpha\over
r_c^{2-\alpha}}\right)=\rho_m(1+z)^3+\rho_r(1+z)^4,
\end{equation}
where $\alpha$ is a phenomenological parameter, and
$r_c=(1-\Omega_m-\Omega_r)^{1/(\alpha-2)}H_0^{-1}$. According to
this Friedmann equation, $E(z)$ is determined by the following
equation:
\begin{equation}
E(z)^2=\Omega_m(1+z)^3+\Omega_{r}(1+z)^4+E(z)^\alpha(1-\Omega_m-\Omega_{r}).
\end{equation} So, this model is a two-parameter model, with the
independent model parameters ${\bm\theta}=\{\Omega_m,~\alpha\}$.
Note that $\alpha=1$ corresponds to the DGP model and $\alpha=0$
corresponds to the cosmological constant model.

\begin{figure}
\includegraphics[scale=0.3, angle=0]{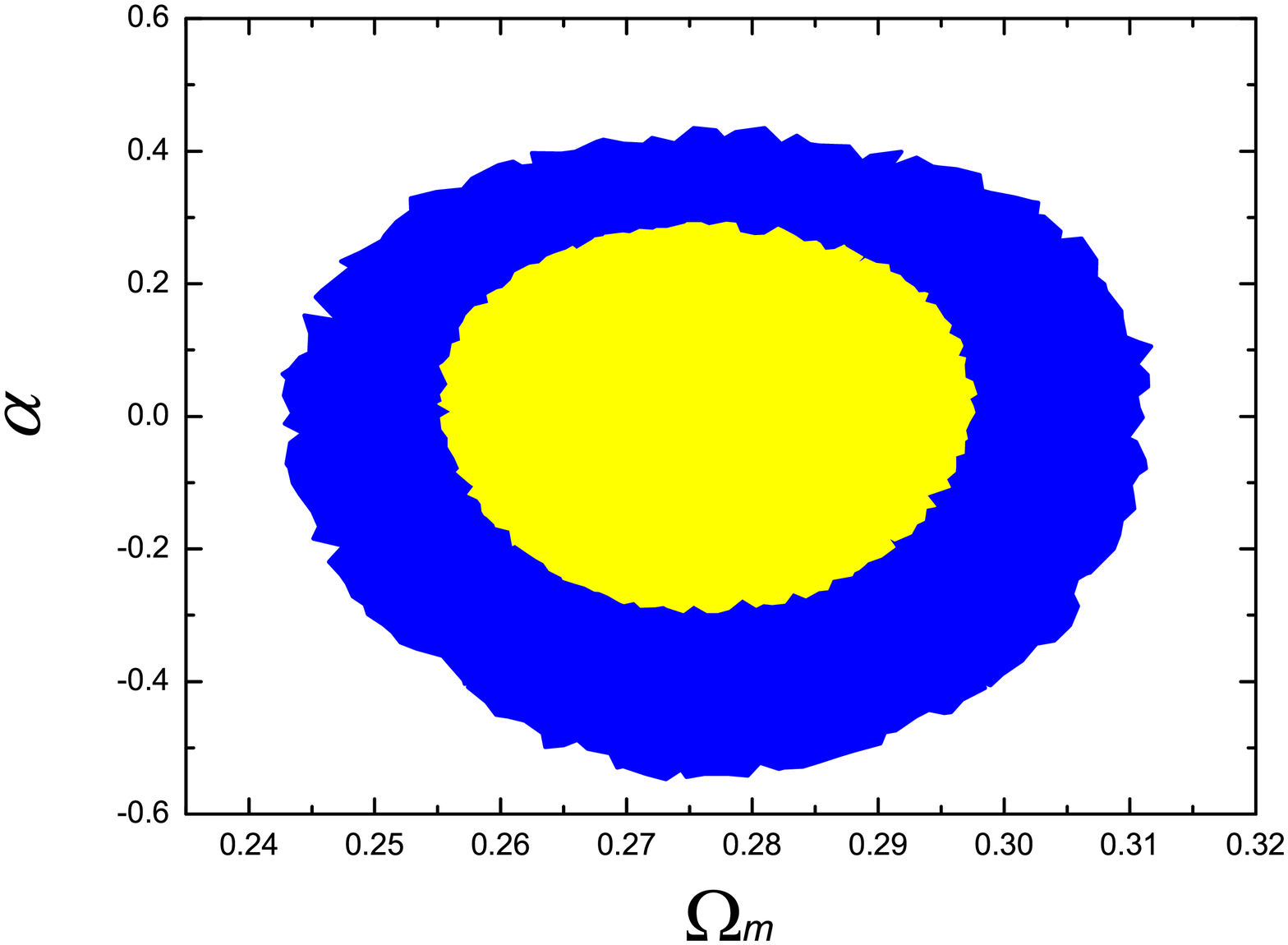}
\includegraphics[scale=0.3, angle=0]{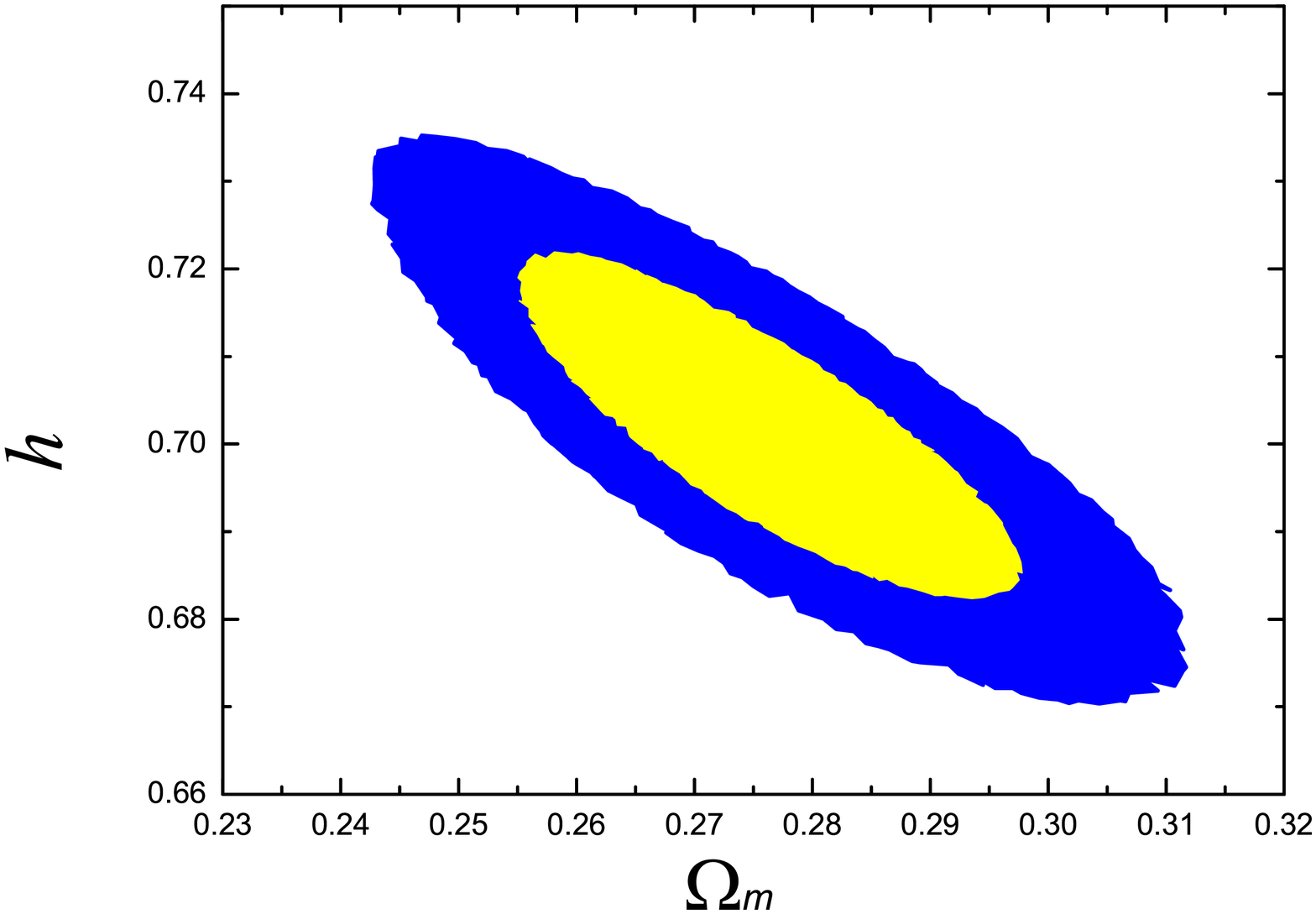}
\caption{The $\alpha$ dark energy model (phenomenological extension
of the DGP model): likelihood contours at 68.3\% and 95.4\%
confidence levels in the $\Omega_m-\alpha$ and $\Omega_m-h$ planes.}
\label{figaDGP}
\end{figure}

Our joint analysis shows that for the $\alpha$DE model the best-fit
parameters and the corresponding $\chi^2_{min}$ are:
\begin{equation}
\Omega_{m}=0.276,\ \ \ \ \alpha=0.030,\ \ \ \ \ h=0.702,\ \ \ \
\chi^2_{min}=468.452.
\end{equation}
We plot the likelihood contours for the $\alpha$DE model in the
$\Omega_m-\alpha$ and $\Omega_m-h$ planes in Fig.~\ref{figaDGP}. We
notice that the best-fit value of $\alpha$ deviates one evidently,
implying that the DGP model is incompatible with the current
observational data. The cosmological constant limit, $\alpha=0$, is
consistent with this model within 1$\sigma$ range. Moreover, we find
that the $\alpha$DE model gives the $\chi^2_{min}$ smaller than the
$\Lambda$ model under our investigation, and its information
criteria results, $\Delta{\rm AIC}=1.991$ and $\Delta{\rm
BIC}=5.987$, also indicate that the $\alpha$DE model fares the best,
except for the $\Lambda$CDM and $w$ models, under the current
observational tests.

\section{Discussion and Conclusion}\label{sec:concl}

In this work, we have considered nine popular dark-energy
cosmological models and tested them against the latest cosmological
data. This includes observational data of SNe Ia from the
Constitution compilation, BAO from the SDSS, the CMB ``WMAP distance
priors'' from the WMAP seven-year observations, and the measurement
of $H_0$ from the HST. We have used the ``strong inflation prior''
that imposes a flatness prior, and explored dark energy models in
the context of such a flat universe assumption. To assess the
various competing dark energy models and make a comparison, we have
applied the information criteria, both the BIC and AIC, in this
analysis.

\begin{figure}
\includegraphics[scale=0.36, angle=0]{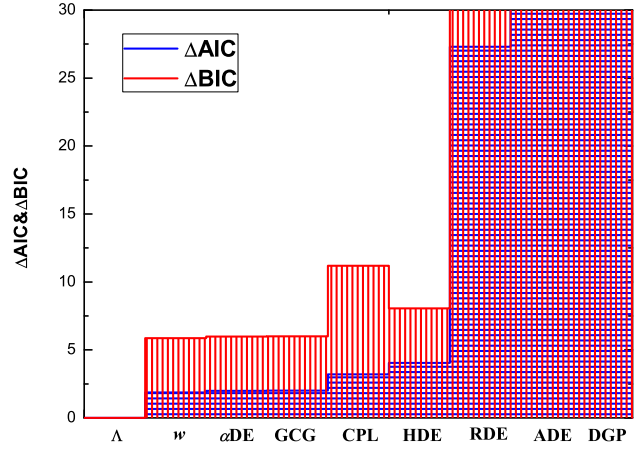}
\caption{Graphical representation of the results in Table
\ref{tab:result}: the values of $\Delta$AIC and $\Delta$BIC for each
model. The order of models from left to right is the same as the
order in Table \ref{tab:result}, which is listed in order of
increasing $\Delta$AIC.} \label{AICBIC}
\end{figure}



For each model, we have outlined the basic equations governing the
evolution of the universe, calculated the best-fit values of its
parameters, and found its $\Delta{\rm AIC}$ and $\Delta{\rm BIC}$
values. Table~\ref{tab:model} summarizes all the models under
consideration and the parameters that describe each model. We have
plotted the likelihood contours of parameters for all the models.
The fit and information criteria results have been summarized in
Table~\ref{tab:result}. Note that since the cosmological constant
model has the lowest values of both AIC and BIC, the values of
$\Delta{\rm AIC}$ and $\Delta{\rm BIC}$ are measured with respect to
this model.

From Table~\ref{tab:model}, we see that according to the number of
parameters the models can be divided into three classes: the
one-parameter models, including $\Lambda$, ADE, and DGP models; the
two-parameter models, including $w$, GCG, HDE, RDE, and $\alpha$DE
models; and the three-parameter model, namely, the CPL model. If we
only compare $\chi^2_{min}$, we find that the $\Lambda$ model is not
the best one, and there are four models, namely, $\alpha$DE, $w$,
GCG, and CPL models, a little bit better than the $\Lambda$ model
according to this criterion. However, if the economical efficiency
is considered, one will find that the $\Lambda$ model is the best
one to explain the current data, since both the AIC and BIC values
it yields are the smallest. Although the CPL model can fit the
current data well and has the lowest $\chi^2_{min}$, it is too
complex (it has three free model parameters) to be necessary, yet.
In the two-parameter models, the $w$ model performs best in
explaining the current data.

We can also classify these models in another way that whether the
model can reduce to the $\Lambda$ model. Some models, such as $w$,
CPL, GCG, and $\alpha$DE, can reduce to the $\Lambda$ model, but the
other ones, namely, HDE, ADE, RDE, and DGP, can not. From
Table~\ref{tab:result}, we see that the models nest the $\Lambda$
all perform well, whereas the models that cannot reduce to the
$\Lambda$ perform illy, except for the HDE model. The HDE model has
the ability in explaining the current data nearly as good as
$\alpha$DE, $w$, and GCG that nest the $\Lambda$. Also, we notice
that although the $\alpha$DE, $w$, GCG, and CPL models can fit the
data well, they actually all tend to collapse to the $\Lambda$ model
with their best-fit parameters.

Out of all the candidate models we consider, it is obvious that the
simplest $\Lambda$ model remains the best one. Following it are a
series of models that give comparably good fits but have more free
parameters. These include the $\alpha$DE, $w$, GCG, and HDE models,
which have two free parameters; and the CPL model, which has three
free parameters. We have shown that the $\alpha$DE, $w$, GCG, and
CPL models can reduce to the $\Lambda$ model and their best-fit
parameters indeed do so (to within 1$\sigma$ ranges). The HDE model
is the sole one that can give a good fit but does not nest
$\Lambda$. The last three models in Table~\ref{tab:result}, RDE,
ADE, and DGP, are clearly disfavored. They have fewer parameters
than models like CPL, but they score poorly because they are unable
to provide a good fit to the data. They cannot reduce to the
$\Lambda$ model for any values of their parameters. We provide a
graphical representation of the IC results in Fig.~\ref{AICBIC}
which directly shows the scores (in the AIC and BIC tests) the
models gain.

In conclusion, given the current quality of the observational data,
and with the assumption of a flat universe, information criteria
indicate that the cosmological constant model is still the best one
and there is no reason to prefer any more complex model. This
conclusion is in accordance with the previous work by Davis et al.
\cite{Davis:2007na}. We look forward to seeing whether this
conclusion could be changed by future more accurate data.

\begin{acknowledgments}
We would like to thank Yun-Gui Gong, Shuang Wang and Tower Wang for
helpful discussions and suggestions. This work was supported by the
Natural Science Foundation of China under Grant Nos. 10705041,
10821504, 10975032 and 10975172, and Ministry of Science and
Technology 973 program under Grant No. 2007CB815401.
\end{acknowledgments}


\end{document}